\newif\ifFULL
\FULLtrue
\documentclass[conference]{IEEEtran}
\IEEEoverridecommandlockouts
\usepackage{cite}
\usepackage{amsmath,amssymb,amsfonts}
\usepackage{mathrsfs}
\usepackage{amsthm}
\usepackage{algorithm}
\usepackage[noend]{algpseudocode}
\usepackage{multicol}
\usepackage{stmaryrd}
\usepackage{graphicx}
\usepackage{textcomp}
\usepackage{xcolor}
\usepackage{changes}
\usepackage{ulem}
\usepackage{bbm}
\usepackage{lipsum,adjustbox}
 \usepackage{pgfplots}

\usepackage{hyperref}
\hypersetup{
 colorlinks,
 linkcolor={blue!100!black},
 citecolor={blue!100!black},
 urlcolor={blue!80!black}
}
\usepackage[numbers,sort&compress]{natbib}
\newtheorem{theorem}{Theorem}

\newtheorem{lemma}{Lemma}

\theoremstyle{definition}
\newtheorem{definition}{Definition}
\newtheorem{remark}{Remark}

\usepackage{soul} 

\newcommand{\etal}{\textit{et al. }}

\newcommand{\coded}[1]{\widetilde{#1}}

\newcommand{\ceil}[1]{ {\lceil#1\rceil}}

\algblockdefx{As}{EndAs}[1]{\textbf{as} #1}

\newcommand{\bbF}{\mathbb{F}}

\newcommand{\bfg}{\mathbf{g}}

\newcommand{\bfs}{\mathbf{s}}

\newcommand{\bfx}{\mathbf{x}}

\newcommand{\bfA}{\mathbf{A}}

\newcommand{\cR}{\mathcal{R}}
\newcommand{\cS}{\mathcal{S}}
\newcommand{\cT}{\mathcal{T}}

\newcommand{\pt}{{(t)}}

\DeclareMathOperator {\diag}{diag}

\begin{document}
\title{All-to-All Encode in Synchronous Systems}

\author{\textbf{Canran Wang} and  \IEEEauthorblockN{\textbf{Netanel Raviv}}
	\IEEEauthorblockA{
		Department of Computer Science and Engineering, Washington University in St. Louis, St. Louis 63130, MO, USA\\
		}}

\maketitle
\thispagestyle{plain}
\pagestyle{plain}

\begin{abstract}

We define~\emph{all-to-all encode}, a collective communication operation serving as a primitive in decentralized computation and storage systems. 
Consider a scenario where every processor initially has a data packet and requires a linear combination of all data packets; the linear combinations are distinct from one processor to another, and are specified by a generator matrix of an error correcting code.
We use a linear network model, in which processors transmit linear combinations of their data and previously received packets, and adopt a standard synchronous system setting to analyze its communication cost. We provide a universal algorithm which computes any matrix in this model by only varying intermediate coefficients, and prove its optimality. 
When the generator matrix is of the Vandermonde or Lagrange type, we further optimize the communication efficiency of the proposed algorithm. 
\end{abstract}

\section{Introduction}
The interest in coding for decentralized systems has increased lately, due to emerging applications in blockchains~\cite{Polyshard,MyPaper,Discrepancy}, sensor networks~\cite{Wireless}, and the internet of things~\cite{IOT}. 
In such systems, raw data is generated independently in distributed source nodes, and then encoded and delivered to distributed sink nodes, without a central authority which orchestrates the operation. 
Examples include encoding for reliable distributed storage (e.g., with a Reed-Solomon or a random code~\cite{Dimakis}) or for straggler-resilient distributed computation~\cite{LCC}.
To study the communication cost of this setting, we focus on the following fundamental collective communication operation.

\begin{definition}
\textbf{(All-to-all encode)}
Consider a distributed system with~$K$ processors and no master processor. Every processor~$k\in[0,K-1]\triangleq\{ 0,1,\ldots,K-1\} $ initially possess an~\emph{initial packet}~$x_k\in\bbF_q$, where~$\bbF_q$ is a finite field with~$q$ elements, and obtains a coded packet~$\coded{x}_k\in\bbF_q$ after the communication operation.
For~$A\in\bbF_q^{K\times K}$ that is known a priori to all processors, the coded packets are defined as
$$
(\coded{x}_0,\ldots,\coded{x}_{K-1}) = ({x_0},\ldots,{x_{K-1}})\cdot A.
$$
\end{definition}

That is, in an all-to-all encode operation, every node has its own data packet, and wishes to obtain a linear combination of all other packets in the system. 
An algorithm which successfully achieves all-to-all encode for a given~$A$ and every~$x_0,\ldots,x_{K-1}\in\bbF_q$ is said to \textit{compute~$A$}. 
Computing matrices in this context emerges during the encoding phase in coded decentralized systems, as described shortly.

We adopt the popular communication model of~\cite{A2A}, in which the processors are connected by a synchronous network, and messages can pass between any two of them.
The system operates in consecutive communication rounds, during each a processor can simultaneously send and receive~$1$ message, which might contain multiple field elements, through each one of its~$p<K$ ports.
Inspired by the well-familiar network-coding literature~\cite{NetworkInformationFlow,LinearNetworkCoding}, we adopt a linear network model in which processors transmit linear combinations of their own data and previously received packets.

Similar to~\cite{A2A}, to capture the communication cost we consider the time to pass a message containing~$d$ field elements as~$\beta+d\cdot \tau$, where~$\beta$ and~$\tau$ are system parameters;~$\beta$ is the startup time of each message delivery, and~$\tau$ refers to the per-element cost. We focus on two measures of communication:
\begin{itemize}
    \item $C_1$: the number of rounds incurred by the algorithm.
    \item $C_2$: the total number of field elements transferred in a sequence during the operation. That is,~$C_2=\sum_{t\in[T]}d_t$ in a~$T$-round algorithm, where~$d_t$ is the size of message containing the largest number of field elements\footnote{Since the operation proceeds in rounds, the largest message in an individual round determines the time duration of this round.} transferred in round~$t$, among all ports of all processors.
\end{itemize}
Our main goal in this paper is optimizing the total communication cost, given by~$C_1\cdot\beta+C_2\cdot\tau$. 
Note that since each round includes sending at least one element from one node to another, it follows that~$C_2 \geq C_1$.

Clearly, all-to-all encode strongly depends on the properties of the underlying matrix~$A$. 
Moreover, any algorithm for this problem contains two separate components, \textit{scheduling} and \textit{coding scheme}. 
The former determines which processor communicates with which other processors at each round, and the latter determines the coefficients in the linear combinations that processors transmit to one another. 
Motivated by this distinction, we address the all-to-all encode problem on two levels, the \textit{universal} and the \textit{specific}. 

\begin{figure}
    \centering
\tikzset{every picture/.style={line width=0.75pt}} 
\begin{tikzpicture}[x=0.5pt,y=0.5pt,yscale=-1,xscale=1]

\draw   (210,168) .. controls (210,163.58) and (213.58,160) .. (218,160) -- (242,160) .. controls (246.42,160) and (250,163.58) .. (250,168) -- (250,192) .. controls (250,196.42) and (246.42,200) .. (242,200) -- (218,200) .. controls (213.58,200) and (210,196.42) .. (210,192) -- cycle ;
\draw   (210,128) .. controls (210,123.58) and (213.58,120) .. (218,120) -- (242,120) .. controls (246.42,120) and (250,123.58) .. (250,128) -- (250,152) .. controls (250,156.42) and (246.42,160) .. (242,160) -- (218,160) .. controls (213.58,160) and (210,156.42) .. (210,152) -- cycle ;
\draw   (210,88) .. controls (210,83.58) and (213.58,80) .. (218,80) -- (242,80) .. controls (246.42,80) and (250,83.58) .. (250,88) -- (250,112) .. controls (250,116.42) and (246.42,120) .. (242,120) -- (218,120) .. controls (213.58,120) and (210,116.42) .. (210,112) -- cycle ;
\draw   (210,48) .. controls (210,43.58) and (213.58,40) .. (218,40) -- (242,40) .. controls (246.42,40) and (250,43.58) .. (250,48) -- (250,72) .. controls (250,76.42) and (246.42,80) .. (242,80) -- (218,80) .. controls (213.58,80) and (210,76.42) .. (210,72) -- cycle ;
\draw   (90,88) .. controls (90,83.58) and (93.58,80) .. (98,80) -- (122,80) .. controls (126.42,80) and (130,83.58) .. (130,88) -- (130,112) .. controls (130,116.42) and (126.42,120) .. (122,120) -- (98,120) .. controls (93.58,120) and (90,116.42) .. (90,112) -- cycle ;
\draw   (90,168) .. controls (90,163.58) and (93.58,160) .. (98,160) -- (122,160) .. controls (126.42,160) and (130,163.58) .. (130,168) -- (130,192) .. controls (130,196.42) and (126.42,200) .. (122,200) -- (98,200) .. controls (93.58,200) and (90,196.42) .. (90,192) -- cycle ;
\draw   (90,128) .. controls (90,123.58) and (93.58,120) .. (98,120) -- (122,120) .. controls (126.42,120) and (130,123.58) .. (130,128) -- (130,152) .. controls (130,156.42) and (126.42,160) .. (122,160) -- (98,160) .. controls (93.58,160) and (90,156.42) .. (90,152) -- cycle ;
\draw   (90,48) .. controls (90,43.58) and (93.58,40) .. (98,40) -- (122,40) .. controls (126.42,40) and (130,43.58) .. (130,48) -- (130,72) .. controls (130,76.42) and (126.42,80) .. (122,80) -- (98,80) .. controls (93.58,80) and (90,76.42) .. (90,72) -- cycle ;
\draw   (410,168) .. controls (410,163.58) and (413.58,160) .. (418,160) -- (442,160) .. controls (446.42,160) and (450,163.58) .. (450,168) -- (450,192) .. controls (450,196.42) and (446.42,200) .. (442,200) -- (418,200) .. controls (413.58,200) and (410,196.42) .. (410,192) -- cycle ;
\draw   (410,128) .. controls (410,123.58) and (413.58,120) .. (418,120) -- (442,120) .. controls (446.42,120) and (450,123.58) .. (450,128) -- (450,152) .. controls (450,156.42) and (446.42,160) .. (442,160) -- (418,160) .. controls (413.58,160) and (410,156.42) .. (410,152) -- cycle ;
\draw   (410,88) .. controls (410,83.58) and (413.58,80) .. (418,80) -- (442,80) .. controls (446.42,80) and (450,83.58) .. (450,88) -- (450,112) .. controls (450,116.42) and (446.42,120) .. (442,120) -- (418,120) .. controls (413.58,120) and (410,116.42) .. (410,112) -- cycle ;
\draw   (410,48) .. controls (410,43.58) and (413.58,40) .. (418,40) -- (442,40) .. controls (446.42,40) and (450,43.58) .. (450,48) -- (450,72) .. controls (450,76.42) and (446.42,80) .. (442,80) -- (418,80) .. controls (413.58,80) and (410,76.42) .. (410,72) -- cycle ;
\draw [color={rgb, 255:red, 208; green, 2; blue, 27 }  ,draw opacity=1 ]   (130,60) .. controls (160.18,60.34) and (164.63,98.29) .. (207.34,99.94) ;
\draw [shift={(210,100)}, rotate = 180.19] [fill={rgb, 255:red, 208; green, 2; blue, 27 }  ,fill opacity=1 ][line width=0.08]  [draw opacity=0] (8.93,-4.29) -- (0,0) -- (8.93,4.29) -- cycle    ;
\draw [color={rgb, 255:red, 248; green, 231; blue, 28 }  ,draw opacity=1 ]   (130,100) .. controls (160.18,100.34) and (164.63,138.29) .. (207.34,139.94) ;
\draw [shift={(210,140)}, rotate = 180.19] [fill={rgb, 255:red, 248; green, 231; blue, 28 }  ,fill opacity=1 ][line width=0.08]  [draw opacity=0] (8.93,-4.29) -- (0,0) -- (8.93,4.29) -- cycle    ;
\draw [color={rgb, 255:red, 144; green, 19; blue, 254 }  ,draw opacity=1 ]   (130,140) .. controls (160.18,140.34) and (164.63,178.29) .. (207.34,179.94) ;
\draw [shift={(210,180)}, rotate = 180.19] [fill={rgb, 255:red, 144; green, 19; blue, 254 }  ,fill opacity=1 ][line width=0.08]  [draw opacity=0] (8.93,-4.29) -- (0,0) -- (8.93,4.29) -- cycle    ;
\draw [color={rgb, 255:red, 0; green, 0; blue, 0 }  ,draw opacity=1 ]   (130,170) .. controls (212.86,170.35) and (131.46,62.05) .. (207.64,60.03) ;
\draw [shift={(210,60)}, rotate = 180.11] [fill={rgb, 255:red, 0; green, 0; blue, 0 }  ,fill opacity=1 ][line width=0.08]  [draw opacity=0] (8.93,-4.29) -- (0,0) -- (8.93,4.29) -- cycle    ;
\draw [color={rgb, 255:red, 248; green, 231; blue, 28 }  ,draw opacity=1 ]   (250,100) .. controls (349.29,100.34) and (349.81,177.97) .. (407.33,179.96) ;
\draw [shift={(410,180)}, rotate = 179.67] [fill={rgb, 255:red, 248; green, 231; blue, 28 }  ,fill opacity=1 ][line width=0.08]  [draw opacity=0] (8.93,-4.29) -- (0,0) -- (8.93,4.29) -- cycle    ;
\draw [color={rgb, 255:red, 208; green, 2; blue, 27 }  ,draw opacity=1 ]   (250,60) .. controls (337.96,60.34) and (371.3,99.16) .. (408.31,138.22) ;
\draw [shift={(410,140)}, rotate = 226.44] [fill={rgb, 255:red, 208; green, 2; blue, 27 }  ,fill opacity=1 ][line width=0.08]  [draw opacity=0] (8.93,-4.29) -- (0,0) -- (8.93,4.29) -- cycle    ;
\draw [color={rgb, 255:red, 144; green, 19; blue, 254 }  ,draw opacity=1 ]   (250,140) .. controls (349.29,140.34) and (350.29,63.22) .. (407.35,60.09) ;
\draw [shift={(410,60)}, rotate = 179.18] [fill={rgb, 255:red, 144; green, 19; blue, 254 }  ,fill opacity=1 ][line width=0.08]  [draw opacity=0] (8.93,-4.29) -- (0,0) -- (8.93,4.29) -- cycle    ;
\draw    (250,180) .. controls (347.31,180.34) and (381.43,132.81) .. (408.36,101.87) ;
\draw [shift={(410,100)}, rotate = 131.4] [fill={rgb, 255:red, 0; green, 0; blue, 0 }  ][line width=0.08]  [draw opacity=0] (8.93,-4.29) -- (0,0) -- (8.93,4.29) -- cycle    ;
\draw [color={rgb, 255:red, 208; green, 2; blue, 27 }  ,draw opacity=1 ]   (450,48) .. controls (497.24,29.72) and (493.86,92.74) .. (452.57,73.3) ;
\draw [shift={(450,72)}, rotate = 28.12] [fill={rgb, 255:red, 208; green, 2; blue, 27 }  ,fill opacity=1 ][line width=0.08]  [draw opacity=0] (8.93,-4.29) -- (0,0) -- (8.93,4.29) -- cycle    ;
\draw [color={rgb, 255:red, 248; green, 231; blue, 28 }  ,draw opacity=1 ]   (450,88) .. controls (497.24,69.72) and (493.86,132.74) .. (452.57,113.3) ;
\draw [shift={(450,112)}, rotate = 28.12] [fill={rgb, 255:red, 248; green, 231; blue, 28 }  ,fill opacity=1 ][line width=0.08]  [draw opacity=0] (8.93,-4.29) -- (0,0) -- (8.93,4.29) -- cycle    ;
\draw [color={rgb, 255:red, 144; green, 19; blue, 254 }  ,draw opacity=1 ]   (450,128) .. controls (497.24,109.72) and (493.86,172.74) .. (452.57,153.3) ;
\draw [shift={(450,152)}, rotate = 28.12] [fill={rgb, 255:red, 144; green, 19; blue, 254 }  ,fill opacity=1 ][line width=0.08]  [draw opacity=0] (8.93,-4.29) -- (0,0) -- (8.93,4.29) -- cycle    ;
\draw [color={rgb, 255:red, 0; green, 0; blue, 0 }  ,draw opacity=1 ]   (450,170) .. controls (497.24,151.72) and (493.86,214.74) .. (452.57,195.3) ;
\draw [shift={(450,194)}, rotate = 28.12] [fill={rgb, 255:red, 0; green, 0; blue, 0 }  ,fill opacity=1 ][line width=0.08]  [draw opacity=0] (8.93,-4.29) -- (0,0) -- (8.93,4.29) -- cycle    ;

\draw (105,54) node [anchor=north west][inner sep=0.75pt]  [xscale=0.74,yscale=0.74] [align=left] {$\displaystyle 0$};
\draw (105,93) node [anchor=north west][inner sep=0.75pt]  [xscale=0.74,yscale=0.74] [align=left] {$\displaystyle 1$};
\draw (105,133) node [anchor=north west][inner sep=0.75pt]  [xscale=0.74,yscale=0.74] [align=left] {$\displaystyle 2$};
\draw (105,173) node [anchor=north west][inner sep=0.75pt]  [xscale=0.74,yscale=0.74] [align=left] {$\displaystyle 3$};
\draw (225,54) node [anchor=north west][inner sep=0.75pt]  [xscale=0.74,yscale=0.74] [align=left] {$\displaystyle 0$};
\draw (225,93) node [anchor=north west][inner sep=0.75pt]  [xscale=0.74,yscale=0.74] [align=left] {$\displaystyle 1$};
\draw (225,133) node [anchor=north west][inner sep=0.75pt]  [xscale=0.74,yscale=0.74] [align=left] {$\displaystyle 2$};
\draw (225,173) node [anchor=north west][inner sep=0.75pt]  [xscale=0.74,yscale=0.74] [align=left] {$\displaystyle 3$};
\draw (425,54) node [anchor=north west][inner sep=0.75pt]  [xscale=0.74,yscale=0.74] [align=left] {$\displaystyle 0$};
\draw (425,93) node [anchor=north west][inner sep=0.75pt]  [xscale=0.74,yscale=0.74] [align=left] {$\displaystyle 1$};
\draw (425,133) node [anchor=north west][inner sep=0.75pt]  [xscale=0.74,yscale=0.74] [align=left] {$\displaystyle 2$};
\draw (425,173) node [anchor=north west][inner sep=0.75pt]  [xscale=0.74,yscale=0.74] [align=left] {$\displaystyle 3$};
\draw (131,42) node [anchor=north west][inner sep=0.75pt]  [xscale=0.74,yscale=0.74] [align=left] {$\displaystyle a_{01} x_{0}$};
\draw (131,80) node [anchor=north west][inner sep=0.75pt]  [xscale=0.74,yscale=0.74] [align=left] {$\displaystyle a_{12} x_{1}$};
\draw (131,120) node [anchor=north west][inner sep=0.75pt]  [xscale=0.74,yscale=0.74] [align=left] {$\displaystyle a_{23} x_{2}$};
\draw (132,173) node [anchor=north west][inner sep=0.75pt]  [xscale=0.74,yscale=0.74] [align=left] {$\displaystyle a_{30} x_{3}$};
\draw (251,40) node [anchor=north west][inner sep=0.75pt]  [xscale=0.74,yscale=0.74] [align=left] {$\displaystyle a_{02} x_{0} +a_{32} x_{3}$};
\draw (251,80) node [anchor=north west][inner sep=0.75pt]  [xscale=0.74,yscale=0.74] [align=left] {$\displaystyle a_{13} x_{1} +a_{03} x_{0}$};
\draw (251,140) node [anchor=north west][inner sep=0.75pt]  [xscale=0.74,yscale=0.74] [align=left] {$\displaystyle a_{20} x_{2} +a_{10} x_{1}$};
\draw (251,180) node [anchor=north west][inner sep=0.75pt]  [xscale=0.74,yscale=0.74] [align=left] {$\displaystyle a_{31} x_{3} +a_{21} x_{2}$};

\draw (300,205) node [anchor=north west][inner sep=0.75pt]  [font=\large,xscale=0.75,yscale=0.75] [align=left] {Round 2};
\draw (140,205) node [anchor=north west][inner sep=0.75pt]  [font=\large,xscale=0.75,yscale=0.75] [align=left] {Round 1};

\draw (484,60) node [anchor=north west][inner sep=0.75pt]  [xscale=0.74,yscale=0.74] [align=left] {$\displaystyle +a_{00} x_{0} +a_{30} x_{3}$};
\draw (484,100) node [anchor=north west][inner sep=0.75pt]  [xscale=0.74,yscale=0.74] [align=left] {$\displaystyle +a_{11} x_{1} +a_{01} x_{0}$};
\draw (483,140) node [anchor=north west][inner sep=0.75pt]  [xscale=0.74,yscale=0.74] [align=left] {$\displaystyle +a_{2}{}_{2} x_{2} +a_{12} x_{1}$};
\draw (484,180) node [anchor=north west][inner sep=0.75pt]  [xscale=0.74,yscale=0.74] [align=left] {$\displaystyle +a_{33} x_{3} +a_{23} x_{2}$};
\draw (484,42) node [anchor=north west][inner sep=0.75pt]  [xscale=0.74,yscale=0.74] [align=left] {$\displaystyle ( a_{20} x_{2} +a_{10} x_{1})$};
\draw (484,82) node [anchor=north west][inner sep=0.75pt]  [xscale=0.74,yscale=0.74] [align=left] {$\displaystyle ( a_{31} x_{3} +a_{21} x_{2})$};
\draw (484,122) node [anchor=north west][inner sep=0.75pt]  [xscale=0.74,yscale=0.74] [align=left] {$\displaystyle ( a_{02} x_{0} +a_{32} x_{3})$};
\draw (484,162) node [anchor=north west][inner sep=0.75pt]  [xscale=0.74,yscale=0.74] [align=left] {$\displaystyle ( a_{13} x_{1} +a_{03} x_{0})$};
\end{tikzpicture}
    \caption{An example of universal algorithm for computing any~$A\in\bbF_q^{4\times 4}$ in~$2$ rounds, when~$p=1$. Every processor~$k\in[0,3]$ wants~$\coded{x}_k=a_{0k}x_0+a_{1k}x_1+a_{2k}x_2+a_{3k}x_3$.
    In the first round, it receives~$a_{(k-1)k}x_{k-1}$ from processor~$k-1$.
    In the second round, it receives~$a_{(k-2)k}x_{k-2}+a_{(k-3)k}x_{k-3}$ from processor~$k-2$. 
    Finally, processor~$k$ sums up the~$2$ received packets with~$a_{kk}x_k$ and obtains~$\coded{x}_k$.}\label{figure:universal}
    \label{fig:universal}
\end{figure}
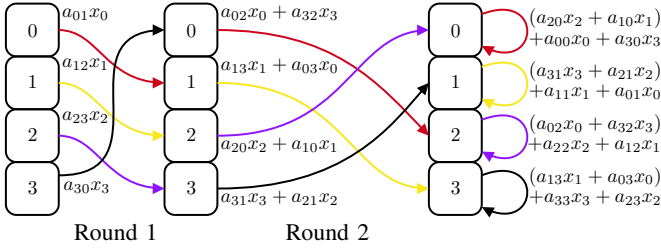

On the universal level, we seek a scheduling by which \textit{any} matrix~$A$ could be computed by only varying the coding scheme, i.e., the coefficients in the transmitted packets throughout the algorithm (see Fig.~\ref{figure:universal}). 
That is, a \textit{universal algorithm} is a series of instructions which indicate which processor communicates with which other processors at each round, alongside a mechanism that for every given~$A$ determines the coefficients that are used by each processor in order to linearly combine previously received packets in each transmission. 
Universal algorithms are important in cases where the scheduling must either be determined prior to knowing~$A$, must apply in several consecutive computations of different matrices, or its simplicity or uniformity are paramount (see Remark~\ref{remark:encoding} below).
Further, universal algorithms can be used as primitives in specific algorithms, as we show in the sequel.

On the specific level, we seek both scheduling and coding scheme that are uniquely tailored towards a specific matrix of interest. 
Clearly, such specific algorithms are important only if they outperform universal ones, since by definition, every universal algorithm subsumes a specific algorithm for all matrices. We are particularly interested in Vandermonde and Lagrange matrices, that are required in Reed-Solomon and Lagrange coded systems, respectively.

Finally, we emphasize that in either the specific or the universal setting, neither the scheduling nor the coding scheme depend on the input packets~$x_i$, but are exclusively determined by the matrix~$A$, that is known to all processors.

\subsection*{Our Contributions}
\begin{itemize}
    \item We provide lower bounds for~$C_1$ and~$C_2$ in a universal algorithm, and propose \textbf{prepare-and-shoot}, a universal algorithm which is optimal in~$C_1$ and achieves the lower bound of~$C_2$ within a factor of~$\sqrt{2}$.
    \item We propose \textbf{draw-and-loose}, a family of specific algorithms for Vandermonde matrices, which optimize the aforementioned universal algorithm in terms of~$C_2$, and provide similar gains for Lagrange matrices.
\end{itemize}

\begin{remark}[All-to-all encode for decentralized encoding]\label{remark:encoding}
Apart from its independent interest, the all-to-all encode operation can be used as a primitive in decentralized coded systems. 
For integers~$N$ and~$K$ such that~$K\vert N$, consider a system with~$N$ processors in which processor~$i\in[K]$ holds~$x_i$, where~$[K]\triangleq\{1,2,\ldots,K\}$. 
Each processor~$j\in[N]$ requires a different linear combination of the~$x_i$'s, defined by the~$j$'th column of a predetermined generator matrix~$G\in\bbF_q^{K\times N}$. 
The all-to-all encode operation defined herein is applicable to this setting as follows. Partition the processors to~$N/K$ subsets of size~$K$ each:~$\{1,\ldots,K\},\{K+1,\ldots,2K\}$, and so on. 
First, each processor~$i\in[K]$ disseminates its~$x_i$ to the processors~$\{\ell K+i\}_{\ell=1}^{N/K-1}$ using a simple tree-structured broadcast protocol with~$\log_{p+1}(N/K)$ rounds. 
Then, each subset runs an all-to-all encode operation to compute the respective~$K\times K$ submatrix of~$G$. 
\end{remark}

\section{Related Works}

Collective communication operations (e.g., one/all-to-all broadcast, one/all-to-all reduce, scatter, gather, etc.) have been studied extensively due to their importance in parallel algorithms, see~\cite{Para} for a thorough introduction to the topic. Yet, to the best of the authors' knowledge, an encompassing treatment of the all-to-all encode operation defined above is conspicuously absent from the literature, and several special cases have been scantly studied in recent years.

Jeong~\etal~\cite{CodedFFT} studies the decentralized encoding process of an~$[N,K]$ \emph{systematic} code as part of a coded FFT algorithm, but did not study the all-to-all encode problem. Decentralized encoding has also been studied by~\cite{Dimakis}, in which processors pass on a \emph{random} linear combination of packets to their neighbors, resulting in an MDS code with high probability. 
Similar problems have been studied in the signal processing literature under the title \emph{Graph Signal Processing}~\cite{GSP}, with a few substantial differences---a graph structure dictates the network connections, communication proceeds in so called graph-shift operations, and computations are over the real or complex fields.
 
\section{Lower Bounds}
We now propose the lower bounds for~$C_1$ and~$C_2$ which apply to any universal algorithm. 

\begin{lemma}\label{lemma:universalC1bound}
Any universal algorithm has~$C_1\geq \ceil{\log_{p+1}K}$.
\end{lemma}
\ifFULL \begin{proof}

Since a universal algorithm must apply for all~$K\times K$ matrices, it must also apply for matrices with no zero entries, i.e., every processor wants a linear combination of all initial packets.
Similar to~\cite[Proposition 2.1]{A2A}, a packet~$x_k$ can reach at most~$(p+1)^t$ processors in the~$p$-port model after round~$t$,. 
Hence, it takes at least~$\ceil{\log_{p+1}K}$ rounds for any packet~$x_k$ to reach all processors.
\end{proof} \fi

\begin{lemma}\label{lemma:universalC2bound}
Any universal algorithm has~$C_2\geq \frac{\sqrt{2K}}{p}-O(1)$.
\end{lemma}
\ifFULL \begin{proof}
We define~\emph{baseline algorithms} as a class of universal algorithms in which every processor passes exactly~$1$ field element through each port during every round; clearly, all baseline algorithms have~$C_1=C_2$.


Given a universal algorithm in which processors transfer at most~$d_t$ packets through one of the ports in round~$t$ (and thus~$C_2=\sum_t d_t$), there exists a corresponding baseline algorithm that ``simulates'' each round~$t$ in the universal algorithm using~$d_t$ rounds, and thus has the same~$C_2=\sum_t d_t$.
Therefore, a universal algorithm cannot outperform all baseline algorithms in terms of~$C_2$. Hence, it suffices to bound~$C_1$, and hence~$C_2$, for baseline algorithms.

We provide a lower bound on~$C_1$ for baseline algorithms by counting possible~\emph{coding schemes}, i.e., the number of ways processors can linearly combine previously received packets during the algorithm. 
Notice that once the scheduling and coding scheme are fixed, an algorithm cannot compute two distinct matrices\footnote{Recall that an algorithm computes a matrix~$A$ only if it computes the product~$xA$ for every~$x=(x_1,\ldots,x_k)\in\bbF_q^K$. 
Hence, if~$xA=xB$ for every~$x\in\bbF_q^K$, then~$\ker(A-B)=\bbF_q^K$, which implies~$A=B$.}. 
Therefore, this value must be greater or equal to~$|\bbF_q^{K\times K}|=q^{K^2}$.

Let~$\bfs_k^\pt$ be the vector of packets received by processor~$k$ prior to the beginning of round~$t\in[T]$.
Clearly, we have~$|\bfs_k^{(t)}|=(t-1)p$.
In round~$t$, for every port~$\rho\in[0,p-1]$, the processor creates and sends a packet~$y_{k,\rho}^{(t)}$ by summing the initial packet~$x_k$ with a linear combination of previously received packets\footnote{Since the receiver will linearly combine the received packets in future rounds of network coding, the initial packet~$x_k$ has coefficient~$1$ to avoid overcounting the coding schemes.}, i.e., 
$$
y_{k,p}^{(t)} = x_k+ (\bfg_k^\pt)^\intercal \cdot\bfs_k^\pt,
$$
where~$\bfg_k^\pt\in\bbF_q^{p(t-1)}$ is a \emph{coding vector} defining the linear combination of elements in~$\bfs_k^\pt$. 
The number of possible coding vectors in round~$t$ is then~$q^{p(t-1)}$. 

Note that there are~$K$ processors in the~$p$-port system, and the baseline algorithm contains~$T$ rounds.
Further, at the end of the~$T$-round baseline algorithm, processor~$k$ obtains~$\coded{x}_k$ by linearly combining the~$Tp+1$ received packets, including the initial packet~$x_k$. 
Therefore, the total number of coding schemes is
\begin{equation}\label{eq:inequality}
    q^{K(Tp+1)}\cdot \prod_{t\in[T]}q^{(t-1)Kp^2}\geq q^{K\cdot K}.
\end{equation}

Taking logarithm on both sides of Equation~\eqref{eq:inequality}, we have
\begin{equation*}
    K(Tp+1)+\sum_{t\in[T]} (t-1)Kp^2 \geq K^2,
\end{equation*}
which simplifies to~$p^2T^2-p(p-2)T + 2 (1-K) \geq 0$, and hence

\begin{align*}
    C_1=T &\geq \frac{1}{2}-\frac{1}{p}+\sqrt{\frac{1}{4}-\frac{1}{p}-\frac{1}{p^2}+\frac{2K}{p^2}}\\&=\frac{\sqrt{2K}}{p}-O(1).\qedhere
\end{align*}

\end{proof} \fi
\begin{remark}[Lower bound for specific algorithms]\label{remark:bounds4SpecificMatrices}
Clearly, specific algorithms can perform at least as good as universal ones in a any figure of merit; this is since any universal algorithm subsumes specific algorithms for all matrices by definition. 
Providing bound pertaining to specific algorithms proved to be a difficult task, and several such bounds will appear in future versions of this paper. 
Yet, it is readily verified that any matrix which contains a non-zero row cannot be computed with~$C_1 < \log_{p+1}K$; this is due to the simple fact that a given packet~$x_k$ cannot be disseminated to all~$K$ processors in less than this many rounds.  
\end{remark}

\section{Prepare and Shoot:\\An Optimal Universal Algorithm}\label{section:universalAlgorithm}

In this section, we propose a universal algorithm that computes any matrix~$A$. The proposed algorithm consists of two phases, \emph{prepare} ($T_p$ rounds) and~\emph{shoot} ($T_s$ rounds). 
Let~$L$ be the maximum integer such that~$(p+1)^{L}<K$. 
If~$L$ is even let~$T_p=L/2+1$ and~$T_s=L/2$. If~$L$ is odd let~$T_p=T_s=(L+1)/2$. 
In either case, the proposed algorithm has the optimal~$C_1=T_s+T_p=\ceil{\log_{p+1}K}$ (see Lemma~\ref{lemma:universalC1bound}). 
To describe the phases, let~$m=(p+1)^{T_p},n=(p+1)^{T_s}$, and hence~$(n-1)m<K\leq nm$. 
For every~$k\in[0,K-1]$, let 
\begin{align*}
    \cR_k^+&=\{k+\ell\mid \ell\in[0,m-1]\},\\ 
    \cR_k^-&=\{k-\ell\mid \ell\in[0,m-1]\},\\
    \cS^+_k&=\{k+\ell\cdot m\mid \ell\in[0,n-1]\},\mbox{ and}\\
    \cS^-_k&=\{k-\ell\cdot m\mid \ell\in[0,n-1]\}.
\end{align*}
For convenience of notation, subscripts are computed~$\bmod K$.

\textbf{Prepare phase:} This phase consists of~$K$ one-to-$m$ broadcasts happening in parallel; each disseminates~$x_k$ from processor~$k\in[0,K-1]$ to processors in~$\cR^+_k$. 
In round~$t$, for every~$k,r\in[0,K-1]$, processor~$k$ forwards~$x_r$ (if present in its internal storage) to processor 
\ifFULL$$k+\rho\cdot\frac{m}{(p+1)^t}$$\fi~through its~$\rho$-th port, for every~$\rho\in[p]$. 
\ifFULL See Figure~\ref{fig:prepare} for an illustrative example.\fi
\begin{lemma}\label{lemma:prepare}
After~$C_{1,\text{prepare}}=T_p$ rounds, every processor~$k$ has obtained~$x_r$ for every~$r\in\cR^-_k$, with~$C_{2,\text{prepare}}= \frac{(p+1)^{T_p}-1}{p}$.
\end{lemma}
\ifFULL \begin{proof}
Consider a tree defined recursively: Initially, processor~$k$ is added as the root. 
In every subsequent round~$t$, every processor~$r$ in the current tree is connected to processor~$r+\rho\cdot\frac{m}{(p+1)^t}$, for every~$\rho\in[0,p-1]$ (i.e., the processors receiving messages from processor~$r$ in round~$t$). 
The recursion is concluded when~$t=T_p$, after which the tree contains processors in~$\cR_k^+$.

Originally only processor~$k$ holds~$x_k$, and since the algorithm employs every processor to forward every packets it has, it follows that every processor in~$\cR_k^+$ obtains~$x_k$ after~$T_p$ rounds.
Therefore, processor~$k$ has obtained~$x_r$ for every~$r\in\cR^-_k$.
Further, as the number of field elements in a message increase by~$p$-fold after each round, we have
\begin{align*}
C_{2,\text{prepare}}&=\sum_{t=1}^{T_p}(p+1)^{t-1}=\frac{(p+1)^{T_p}-1}{p}.\qedhere
\end{align*}
\end{proof} \fi


\ifFULL
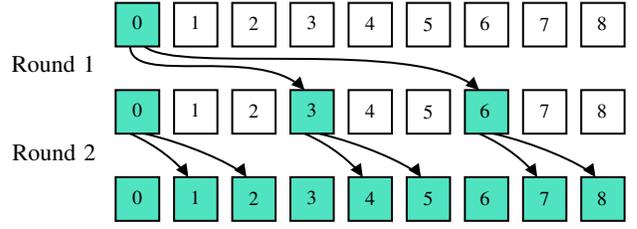
\begin{figure}
    \centering

\tikzset{every picture/.style={line width=0.75pt}} 

\begin{tikzpicture}[x=0.55pt,y=0.55pt,yscale=-1,xscale=1]

\draw  [fill={rgb, 255:red, 80; green, 227; blue, 194 }  ,fill opacity=1 ] (100,90) -- (130,90) -- (130,120) -- (100,120) -- cycle ;
\draw   (140.2,90) -- (170,90) -- (170,120) -- (140.2,120) -- cycle ;
\draw   (180.2,90) -- (210,90) -- (210,120) -- (180.2,120) -- cycle ;
\draw   (220.2,90) -- (250,90) -- (250,120) -- (220.2,120) -- cycle ;
\draw   (260.2,90) -- (290,90) -- (290,120) -- (260.2,120) -- cycle ;
\draw   (300.2,90) -- (330,90) -- (330,120) -- (300.2,120) -- cycle ;
\draw   (340.2,90) -- (370,90) -- (370,120) -- (340.2,120) -- cycle ;
\draw   (380.2,90) -- (410,90) -- (410,120) -- (380.2,120) -- cycle ;
\draw   (420.2,90) -- (450,90) -- (450,120) -- (420.2,120) -- cycle ;
\draw    (110,120) .. controls (110.39,147.65) and (196.07,119.04) .. (228.11,148.15) ;
\draw [shift={(230,150)}, rotate = 226.6] [fill={rgb, 255:red, 0; green, 0; blue, 0 }  ][line width=0.08]  [draw opacity=0] (8.93,-4.29) -- (0,0) -- (8.93,4.29) -- cycle    ;
\draw    (120,120) .. controls (142.26,139.55) and (305.11,113.65) .. (348.12,148.37) ;
\draw [shift={(350,150)}, rotate = 222.96] [fill={rgb, 255:red, 0; green, 0; blue, 0 }  ][line width=0.08]  [draw opacity=0] (8.93,-4.29) -- (0,0) -- (8.93,4.29) -- cycle    ;
\draw    (120,180) .. controls (148.27,186.6) and (179.14,200.03) .. (188.09,207.95) ;
\draw [shift={(190,210)}, rotate = 235.46] [fill={rgb, 255:red, 0; green, 0; blue, 0 }  ][line width=0.08]  [draw opacity=0] (8.93,-4.29) -- (0,0) -- (8.93,4.29) -- cycle    ;
\draw    (110,180) .. controls (129.22,188.32) and (142.51,200.41) .. (148.23,207.58) ;
\draw [shift={(150,210)}, rotate = 236.72] [fill={rgb, 255:red, 0; green, 0; blue, 0 }  ][line width=0.08]  [draw opacity=0] (8.93,-4.29) -- (0,0) -- (8.93,4.29) -- cycle    ;
\draw  [fill={rgb, 255:red, 80; green, 227; blue, 194 }  ,fill opacity=1 ] (100,150) -- (130,150) -- (130,180) -- (100,180) -- cycle ;
\draw   (140.2,150) -- (170,150) -- (170,180) -- (140.2,180) -- cycle ;
\draw   (180.2,150) -- (210,150) -- (210,180) -- (180.2,180) -- cycle ;
\draw  [fill={rgb, 255:red, 80; green, 227; blue, 194 }  ,fill opacity=1 ] (220.2,150) -- (250,150) -- (250,180) -- (220.2,180) -- cycle ;
\draw   (260.2,150) -- (290,150) -- (290,180) -- (260.2,180) -- cycle ;
\draw   (300.2,150) -- (330,150) -- (330,180) -- (300.2,180) -- cycle ;
\draw  [fill={rgb, 255:red, 80; green, 227; blue, 194 }  ,fill opacity=1 ] (340.2,150) -- (370,150) -- (370,180) -- (340.2,180) -- cycle ;
\draw   (380.2,150) -- (410,150) -- (410,180) -- (380.2,180) -- cycle ;
\draw   (420.2,150) -- (450,150) -- (450,180) -- (420.2,180) -- cycle ;
\draw  [fill={rgb, 255:red, 80; green, 227; blue, 194 }  ,fill opacity=1 ] (100,210) -- (130,210) -- (130,240) -- (100,240) -- cycle ;
\draw  [fill={rgb, 255:red, 80; green, 227; blue, 194 }  ,fill opacity=1 ] (140.2,210) -- (170,210) -- (170,240) -- (140.2,240) -- cycle ;
\draw  [fill={rgb, 255:red, 80; green, 227; blue, 194 }  ,fill opacity=1 ] (180.2,210) -- (210,210) -- (210,240) -- (180.2,240) -- cycle ;
\draw  [fill={rgb, 255:red, 80; green, 227; blue, 194 }  ,fill opacity=1 ] (220.2,210) -- (250,210) -- (250,240) -- (220.2,240) -- cycle ;
\draw  [fill={rgb, 255:red, 80; green, 227; blue, 194 }  ,fill opacity=1 ] (260.2,210) -- (290,210) -- (290,240) -- (260.2,240) -- cycle ;
\draw  [fill={rgb, 255:red, 80; green, 227; blue, 194 }  ,fill opacity=1 ] (300.2,210) -- (330,210) -- (330,240) -- (300.2,240) -- cycle ;
\draw  [fill={rgb, 255:red, 80; green, 227; blue, 194 }  ,fill opacity=1 ] (340.2,210) -- (370,210) -- (370,240) -- (340.2,240) -- cycle ;
\draw  [fill={rgb, 255:red, 80; green, 227; blue, 194 }  ,fill opacity=1 ] (380.2,210) -- (410,210) -- (410,240) -- (380.2,240) -- cycle ;
\draw  [fill={rgb, 255:red, 80; green, 227; blue, 194 }  ,fill opacity=1 ] (420.2,210) -- (450,210) -- (450,240) -- (420.2,240) -- cycle ;
\draw    (240,180) .. controls (268.27,186.6) and (299.14,200.03) .. (308.09,207.95) ;
\draw [shift={(310,210)}, rotate = 235.46] [fill={rgb, 255:red, 0; green, 0; blue, 0 }  ][line width=0.08]  [draw opacity=0] (8.93,-4.29) -- (0,0) -- (8.93,4.29) -- cycle    ;
\draw    (230,180) .. controls (249.22,188.32) and (262.51,200.41) .. (268.23,207.58) ;
\draw [shift={(270,210)}, rotate = 236.72] [fill={rgb, 255:red, 0; green, 0; blue, 0 }  ][line width=0.08]  [draw opacity=0] (8.93,-4.29) -- (0,0) -- (8.93,4.29) -- cycle    ;
\draw    (360,180) .. controls (388.27,186.6) and (419.14,200.03) .. (428.09,207.95) ;
\draw [shift={(430,210)}, rotate = 235.46] [fill={rgb, 255:red, 0; green, 0; blue, 0 }  ][line width=0.08]  [draw opacity=0] (8.93,-4.29) -- (0,0) -- (8.93,4.29) -- cycle    ;
\draw    (350,180) .. controls (369.22,188.32) and (382.51,200.41) .. (388.23,207.58) ;
\draw [shift={(390,210)}, rotate = 236.72] [fill={rgb, 255:red, 0; green, 0; blue, 0 }  ][line width=0.08]  [draw opacity=0] (8.93,-4.29) -- (0,0) -- (8.93,4.29) -- cycle    ;

\draw (27,125) node [anchor=north west][inner sep=0.75pt]  [font=\large,xscale=0.75,yscale=0.75] [align=left] {Round 1};
\draw (27.5,186) node [anchor=north west][inner sep=0.75pt]  [font=\large,xscale=0.75,yscale=0.75] [align=left] {Round 2};
\draw (150,97.5) node [anchor=north west][inner sep=0.75pt]  [xscale=0.75,yscale=0.75] [align=left] {1};
\draw (190,98) node [anchor=north west][inner sep=0.75pt]  [xscale=0.75,yscale=0.75] [align=left] {2};
\draw (230,97.5) node [anchor=north west][inner sep=0.75pt]  [xscale=0.75,yscale=0.75] [align=left] {3};
\draw (270,98) node [anchor=north west][inner sep=0.75pt]  [xscale=0.75,yscale=0.75] [align=left] {4};
\draw (310,98.5) node [anchor=north west][inner sep=0.75pt]  [xscale=0.75,yscale=0.75] [align=left] {5};
\draw (349,98) node [anchor=north west][inner sep=0.75pt]  [xscale=0.75,yscale=0.75] [align=left] {6};
\draw (390,98.5) node [anchor=north west][inner sep=0.75pt]  [xscale=0.75,yscale=0.75] [align=left] {7};
\draw (429.5,98) node [anchor=north west][inner sep=0.75pt]  [xscale=0.75,yscale=0.75] [align=left] {8};
\draw (110,97.5) node [anchor=north west][inner sep=0.75pt]  [xscale=0.75,yscale=0.75] [align=left] {0};
\draw (150,157.5) node [anchor=north west][inner sep=0.75pt]  [xscale=0.75,yscale=0.75] [align=left] {1};
\draw (190,158) node [anchor=north west][inner sep=0.75pt]  [xscale=0.75,yscale=0.75] [align=left] {2};
\draw (230,157.5) node [anchor=north west][inner sep=0.75pt]  [xscale=0.75,yscale=0.75] [align=left] {3};
\draw (270,158) node [anchor=north west][inner sep=0.75pt]  [xscale=0.75,yscale=0.75] [align=left] {4};
\draw (310,158.5) node [anchor=north west][inner sep=0.75pt]  [xscale=0.75,yscale=0.75] [align=left] {5};
\draw (349,158) node [anchor=north west][inner sep=0.75pt]  [xscale=0.75,yscale=0.75] [align=left] {6};
\draw (390,158.5) node [anchor=north west][inner sep=0.75pt]  [xscale=0.75,yscale=0.75] [align=left] {7};
\draw (429.5,158) node [anchor=north west][inner sep=0.75pt]  [xscale=0.75,yscale=0.75] [align=left] {8};
\draw (110,157.5) node [anchor=north west][inner sep=0.75pt]  [xscale=0.75,yscale=0.75] [align=left] {0};
\draw (150,217.5) node [anchor=north west][inner sep=0.75pt]  [xscale=0.75,yscale=0.75] [align=left] {1};
\draw (190,218) node [anchor=north west][inner sep=0.75pt]  [xscale=0.75,yscale=0.75] [align=left] {2};
\draw (230,217.5) node [anchor=north west][inner sep=0.75pt]  [xscale=0.75,yscale=0.75] [align=left] {3};
\draw (270,218) node [anchor=north west][inner sep=0.75pt]  [xscale=0.75,yscale=0.75] [align=left] {4};
\draw (310,218.5) node [anchor=north west][inner sep=0.75pt]  [xscale=0.75,yscale=0.75] [align=left] {5};
\draw (349,218) node [anchor=north west][inner sep=0.75pt]  [xscale=0.75,yscale=0.75] [align=left] {6};
\draw (390,218.5) node [anchor=north west][inner sep=0.75pt]  [xscale=0.75,yscale=0.75] [align=left] {7};
\draw (429.5,218) node [anchor=north west][inner sep=0.75pt]  [xscale=0.75,yscale=0.75] [align=left] {8};
\draw (110,217.5) node [anchor=north west][inner sep=0.75pt]  [xscale=0.75,yscale=0.75] [align=left] {0};

\end{tikzpicture}
    \caption{
    Illustration of the dissemination of~$x_0$ in the prepare phase in a~$2$-port system to~$m=9$ processors. 
    In the first round, processor~$0$ broadcasts~$x_0$ to processor~$3$ and processor~$6$. 
    In the second round, processor~$0/3/6$ passes~$x_0$ to processor~$1/4/7$ and processor~$2/5/8$, and concludes the prepare phase.
    }
    \label{fig:prepare}
\end{figure}
\fi

\textbf{Shoot phase:} This phase consists of~$K$ $n$-to-one reduce operations happening in parallel, each intended to communicate the correct linear combination of packets to every processor. 

At the beginning of the phase, every processor~$k$ defines~$n$ variables $w_{k,k},w_{k,k+m},\ldots, w_{k,k+(n-1)m}$ using the information received in the prepare phase (i.e.,~$x_{r}$ for~$r\in\cR_k^-$), and the coefficients of the matrix~$A$. 
Intuitively, the variable $w_{k,s}$ contains a linear combination of $x_r$'s at processor~$k$, whose final destination is processor~$s$; more and more packets will be added to $w_{k,s}$ as the algorithm progresses. 
Specifically, for~$\ell\in[0,n-1]$, initialize~$w_{k,k+\ell\cdot m}$ with~$\bfx_k\cdot A_{k+\ell\cdot m}$, where~$A_{k+\ell\cdot m}$ is the~$(k+\ell\cdot m)$-th column of~$A$, and the non-zero entries of $\bfx_{k}\in\bbF_q^K$ are indexed by elements in~$\cR_k^-$, i.e.,
\begin{equation}\label{eq:reduceTo2}
   \bfx_k[r]=\begin{cases} 
    x_{r} & r\in\cR_k^-\\
    0      & \text{otherwise}
    \end{cases}.
\end{equation}
The goal of this phase is to allow every processor~$k$ to obtain~$y_k=\sum_{s\in\cS_k^-}w_{s,k}=\sum_{s\in\cS_k^-}\bfx_s\cdot\bfA_k$, where~$w_{s,k}$ refers to the content of that variable at the beginning of this phase.

In round~$t\in[1,T_s]$ of the shoot phase, processor~$k$ forwards a message to processor~$k+\rho m^t$, and receives a message from processor~$k-\rho m^t$, for every~$\rho\in[p]$. 
\ifFULL To describe this process, we now define a series of trees for every processor at every round, and an illustrative example can be found in Figure~\ref{fig:reduce}.\fi

\ifFULL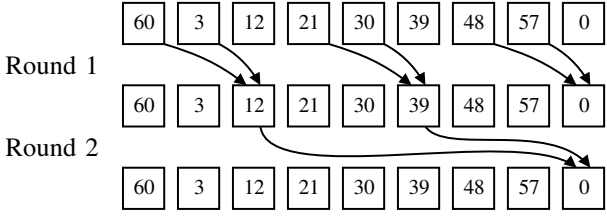
\begin{figure}
    \centering

\tikzset{every picture/.style={line width=0.75pt}} 

\begin{tikzpicture}[x=0.52pt,y=0.52pt,yscale=-1,xscale=1]

\draw   (490,110) -- (520,110) -- (520,140) -- (490,140) -- cycle ;
\draw   (450,110) -- (480,110) -- (480,140) -- (450,140) -- cycle ;
\draw   (410,110) -- (440,110) -- (440,140) -- (410,140) -- cycle ;
\draw   (370,110) -- (400,110) -- (400,140) -- (370,140) -- cycle ;
\draw   (330,110) -- (360,110) -- (360,140) -- (330,140) -- cycle ;
\draw   (290,110) -- (320,110) -- (320,140) -- (290,140) -- cycle ;
\draw   (250,110) -- (280,110) -- (280,140) -- (250,140) -- cycle ;
\draw   (170,110) -- (200,110) -- (200,140) -- (170,140) -- cycle ;
\draw   (210,110) -- (240,110) -- (240,140) -- (210,140) -- cycle ;
\draw    (270,200) .. controls (275.98,246.36) and (455.88,199.19) .. (498.17,228.6) ;
\draw [shift={(500,230)}, rotate = 219.89] [fill={rgb, 255:red, 0; green, 0; blue, 0 }  ][line width=0.08]  [draw opacity=0] (8.93,-4.29) -- (0,0) -- (8.93,4.29) -- cycle    ;
\draw    (390,200) .. controls (404.44,221.3) and (479.64,195.47) .. (508.3,227.94) ;
\draw [shift={(510,230)}, rotate = 232.55] [fill={rgb, 255:red, 0; green, 0; blue, 0 }  ][line width=0.08]  [draw opacity=0] (8.93,-4.29) -- (0,0) -- (8.93,4.29) -- cycle    ;
\draw    (440,140) .. controls (464.38,147.42) and (478.88,153.86) .. (497.63,168.17) ;
\draw [shift={(500,170)}, rotate = 218.05] [fill={rgb, 255:red, 0; green, 0; blue, 0 }  ][line width=0.08]  [draw opacity=0] (8.93,-4.29) -- (0,0) -- (8.93,4.29) -- cycle    ;
\draw    (480,140) .. controls (491.34,146.02) and (500.86,154.62) .. (508.55,167.48) ;
\draw [shift={(510,170)}, rotate = 240.92] [fill={rgb, 255:red, 0; green, 0; blue, 0 }  ][line width=0.08]  [draw opacity=0] (8.93,-4.29) -- (0,0) -- (8.93,4.29) -- cycle    ;
\draw    (320,140) .. controls (344.38,147.42) and (358.88,153.86) .. (377.63,168.17) ;
\draw [shift={(380,170)}, rotate = 218.05] [fill={rgb, 255:red, 0; green, 0; blue, 0 }  ][line width=0.08]  [draw opacity=0] (8.93,-4.29) -- (0,0) -- (8.93,4.29) -- cycle    ;
\draw    (360,140) .. controls (371.34,146.02) and (380.86,154.62) .. (388.55,167.48) ;
\draw [shift={(390,170)}, rotate = 240.92] [fill={rgb, 255:red, 0; green, 0; blue, 0 }  ][line width=0.08]  [draw opacity=0] (8.93,-4.29) -- (0,0) -- (8.93,4.29) -- cycle    ;
\draw    (200,140) .. controls (224.38,147.42) and (238.88,153.86) .. (257.63,168.17) ;
\draw [shift={(260,170)}, rotate = 218.05] [fill={rgb, 255:red, 0; green, 0; blue, 0 }  ][line width=0.08]  [draw opacity=0] (8.93,-4.29) -- (0,0) -- (8.93,4.29) -- cycle    ;
\draw    (240,140) .. controls (251.34,146.02) and (260.86,154.62) .. (268.55,167.48) ;
\draw [shift={(270,170)}, rotate = 240.92] [fill={rgb, 255:red, 0; green, 0; blue, 0 }  ][line width=0.08]  [draw opacity=0] (8.93,-4.29) -- (0,0) -- (8.93,4.29) -- cycle    ;
\draw   (490,170) -- (520,170) -- (520,200) -- (490,200) -- cycle ;
\draw   (450,170) -- (480,170) -- (480,200) -- (450,200) -- cycle ;
\draw   (410,170) -- (440,170) -- (440,200) -- (410,200) -- cycle ;
\draw   (370,170) -- (400,170) -- (400,200) -- (370,200) -- cycle ;
\draw   (330,170) -- (360,170) -- (360,200) -- (330,200) -- cycle ;
\draw   (290,170) -- (320,170) -- (320,200) -- (290,200) -- cycle ;
\draw   (250,170) -- (280,170) -- (280,200) -- (250,200) -- cycle ;
\draw   (170,170) -- (200,170) -- (200,200) -- (170,200) -- cycle ;
\draw   (210,170) -- (240,170) -- (240,200) -- (210,200) -- cycle ;
\draw   (490,230) -- (520,230) -- (520,260) -- (490,260) -- cycle ;
\draw   (450,230) -- (480,230) -- (480,260) -- (450,260) -- cycle ;
\draw   (410,230) -- (440,230) -- (440,260) -- (410,260) -- cycle ;
\draw   (370,230) -- (400,230) -- (400,260) -- (370,260) -- cycle ;
\draw   (330,230) -- (360,230) -- (360,260) -- (330,260) -- cycle ;
\draw   (290,230) -- (320,230) -- (320,260) -- (290,260) -- cycle ;
\draw   (250,230) -- (280,230) -- (280,260) -- (250,260) -- cycle ;
\draw   (170,230) -- (200,230) -- (200,260) -- (170,260) -- cycle ;
\draw   (210,230) -- (240,230) -- (240,260) -- (210,260) -- cycle ;

\draw (499.67,116.67) node [anchor=north west][inner sep=0.75pt]  [font=\footnotesize] [align=left] {0};
\draw (455.67,116.67) node [anchor=north west][inner sep=0.75pt]  [font=\footnotesize] [align=left] {57};
\draw (415.67,116.67) node [anchor=north west][inner sep=0.75pt]  [font=\footnotesize] [align=left] {48};
\draw (375.67,116.67) node [anchor=north west][inner sep=0.75pt]  [font=\footnotesize] [align=left] {39};
\draw (335.67,116.67) node [anchor=north west][inner sep=0.75pt]  [font=\footnotesize] [align=left] {30};
\draw (295.67,116.67) node [anchor=north west][inner sep=0.75pt]  [font=\footnotesize] [align=left] {21};
\draw (255.67,116.67) node [anchor=north west][inner sep=0.75pt]  [font=\footnotesize] [align=left] {12};
\draw (175.67,116.67) node [anchor=north west][inner sep=0.75pt]   [align=left] {{\footnotesize 60}};
\draw (219.67,116.67) node [anchor=north west][inner sep=0.75pt]   [align=left] {{\footnotesize 3}};
\draw (150,155) node   [align=left] {\begin{minipage}[lt]{68pt}\setlength\topsep{0pt}
Round 1
\end{minipage}};
\draw (150,215) node   [align=left] {\begin{minipage}[lt]{68pt}\setlength\topsep{0pt}
Round 2
\end{minipage}};
\draw (499.67,176.67) node [anchor=north west][inner sep=0.75pt]  [font=\footnotesize] [align=left] {0};
\draw (455.67,176.67) node [anchor=north west][inner sep=0.75pt]  [font=\footnotesize] [align=left] {57};
\draw (415.67,176.67) node [anchor=north west][inner sep=0.75pt]  [font=\footnotesize] [align=left] {48};
\draw (375.67,176.67) node [anchor=north west][inner sep=0.75pt]  [font=\footnotesize] [align=left] {39};
\draw (335.67,176.67) node [anchor=north west][inner sep=0.75pt]  [font=\footnotesize] [align=left] {30};
\draw (295.67,176.67) node [anchor=north west][inner sep=0.75pt]  [font=\footnotesize] [align=left] {21};
\draw (255.67,176.67) node [anchor=north west][inner sep=0.75pt]  [font=\footnotesize] [align=left] {12};
\draw (175.67,176.67) node [anchor=north west][inner sep=0.75pt]   [align=left] {{\footnotesize 60}};
\draw (219.67,176.67) node [anchor=north west][inner sep=0.75pt]   [align=left] {{\footnotesize 3}};
\draw (499.67,236.67) node [anchor=north west][inner sep=0.75pt]  [font=\footnotesize] [align=left] {0};
\draw (455.67,236.67) node [anchor=north west][inner sep=0.75pt]  [font=\footnotesize] [align=left] {57};
\draw (415.67,236.67) node [anchor=north west][inner sep=0.75pt]  [font=\footnotesize] [align=left] {48};
\draw (375.67,236.67) node [anchor=north west][inner sep=0.75pt]  [font=\footnotesize] [align=left] {39};
\draw (335.67,236.67) node [anchor=north west][inner sep=0.75pt]  [font=\footnotesize] [align=left] {30};
\draw (295.67,236.67) node [anchor=north west][inner sep=0.75pt]  [font=\footnotesize] [align=left] {21};
\draw (255.67,236.67) node [anchor=north west][inner sep=0.75pt]  [font=\footnotesize] [align=left] {12};
\draw (175.67,236.67) node [anchor=north west][inner sep=0.75pt]   [align=left] {{\footnotesize 60}};
\draw (219.67,236.67) node [anchor=north west][inner sep=0.75pt]   [align=left] {{\footnotesize 3}};

\end{tikzpicture}

    \caption{Illustration of the reduce operation from processors~$60, 3,\ldots,57$ to processor~$0$ in the shoot phase in a~$2$-port system with~$K=65$.}
    \label{fig:reduce}
\end{figure}\fi

For every round~$t$, let~$\cT_k^{(t)}$ be a tree defined recursively in~$T_s-t$ steps, as follows.
Initially, processor~$k$ is added as the root. 
In every subsequent step~$\tau$, for~$\tau\in[1,T_s-t]$, every processor~$r$ in the current tree is connected to processor~$r+\rho m^{t+\tau}$, for every~$\rho\in[p]$. 
In particular, the tree~$\cT_k^{(T_s)}$ contains only the root~$k$. 
Intuitively, the tree~$\cT_k^{(t)}$ contains processors that the processor~$k$ is connecting to, directly or indirectly, in rounds following round~$t$. 

\begin{algorithm}
\caption{Shoot Phase (for processor~$k$)}\label{alg:sp}
\begin{algorithmic}[1]
    \State Initialize~$w_{k,k},w_{k,k+m},\ldots,w_{k,k+(n-1)m}$.
    \For{$t\gets 1,2,\ldots,T_s$}
        \For{$\rho=1$ to~$p$}\Comment{As a sender}
            \State $s_\text{out}=k+\rho m^t$
            \State send~$w_{k,r}$ to processor~$s_\text{out}$ for every~$r\in \cT^{(t)}_{s_\text{out}}$
        \EndFor
        \For{$\rho=1$ to~$p$}\Comment{As a receiver}
            \State $s_\text{in}=k-\rho m^t$
            \For{$r\in \cT^{(t)}_k$}
                \State receive~$w_{s_\text{in},r}$ from processor~$s_\text{in}$
                \State assign~$w_{k,r}\gets w_{k,r}+w_{s_\text{in},r}$
            \EndFor
        \EndFor
    \EndFor
    \State \textbf{Output} $w_{k,k}$ as~$y_k$
\end{algorithmic}
\end{algorithm}

At round~$t$, processor~$k$ sends a message to processor~$s_\text{out}=k+\rho m^\tau$, and receives a message from processor~$s_\text{in}=k-\rho m^\tau$ through the~$\rho$-th port for~$\rho\in[p]$.
The sent message contains~$w_{s_\text{out},r}$ for every~$r\in\cT_{s_\text{out}}^{(t)}$.
The received message contains~$w_{s_\text{in},r}$ for every~$r\in\cT_k^{(t)}$, and the processor~$k$ updates its internal storage, letting~$w_{k,r}\gets w_{k,r}+w_{s_\text{in},r}$. 
The details are given in Algorithm~\ref{alg:sp}, and the correctness is as follows.
\begin{lemma}\label{lemma:shoot}
After~$C_{1,\text{shoot}}=T_s$ rounds, every processor~$k$ has obtained~$y_k=\sum_{r\in\cS_k^-}\bfx_r\cdot\bfA_k$, with~$C_{2,\text{shoot}}= \frac{(p+1)^{T_s}-1}{p}$. 
\end{lemma}
\ifFULL \begin{proof}
We show that the variable~$w_{k,k}$ stores~$y_k$ for every~$k$ at the end of the algorithm.
The proof is based on an recursively defined tree~$\cT_k'$: in step~$\tau=0$ processor~$k$ is added as the root. In every subsequent step~$\tau\ge 1$ every processor~$r$ in the current tree is connected to processor~$r-\rho m^\tau$, for every~$\rho\in[p]$.
That is, during each step~$\tau$, we add processors that \textit{sent} messages to the existing processors at round~$t=T_s-\tau+1$.

Observe that the tree~$\cT_k'$ contains all processors in~$\cS_k^-$ after~$T_s$ steps of the recursion.
For each processor~$r\in\cT_k'$ added in step~$\tau$, processor~$k$ is present in the tree~$\cT_r^{(t)}$.
Hence, at round~$t$, the variable~$w_{r,k}$ is transmitted from processor~$r$ to its parent~$s$ in~$\cT_k'$, and summed with~$w_{s,k}$. 

Traversing from the processors added at step~$T_s$, the packets~$\{\bfx_r\cdot\bfA_k\}_{r\in\cT_k'}$ are summed and transmitted to the root~$k$, and stored in the variable~$w_{k,k}$.
Note that tree~$\cT_k'$ contains exactly the processors in~$\cS_k^-$, and hence~$w_{k,k}=y_k$ stores~$\sum_{r\in\cS_k^-}\bfx_r\cdot\bfA_k$.

Finally, observe that~$|\cT_k^{(t)}|=\frac{n}{(p+1)^t}$, which is the number of field elements sent by processor~$k$ at round~$t$ through each of its ports. Hence, summing over all rounds, we have 
\begin{align*}
    C_{2,\text{shoot}}&=\sum_{t=1}^{T_s}(p+1)^{T_s-t}=\frac{(p+1)^{T_s}-1}{p}.\qedhere
\end{align*}
\end{proof} \fi

Finally, in the most general case where~$K<mn$, some overlap of indices need to be resolved, as some computation results have been summed up twice.
In particular, observe that~$\cR_k^-\cap \cR_{k-(n-1)m}^-=[k-nm+1,k]$, which is an empty set only if~$nm=K$, as we assumed that~$(n-1)m<K\le nm$, and since indices are computed~$\bmod K$.
Therefore,
\begin{equation}\label{eq:xkFromyk}
  y_k=\coded{x}_k+\sum_{r\in[k-mn+1,k]}A_{r,k}x_r,
\end{equation}
from which processor~$k$ can individually compute~$\coded{x}_k$ with no communication, by computing the r.h.s sum and subtracting from~$y_k$.
This concludes the prepare-and-shoot algorithm, and provides the following by Lemma~\ref{lemma:prepare} and Lemma~\ref{lemma:shoot}.
\begin{theorem}
    The prepare-and-shoot algorithm has~$C_1=T_p+T_s=\ceil{\log_{p+1}K}$ and
\begin{equation*}
    C_2= \begin{cases}
    \frac{2(p+1)^{(L+1)/2}-2}{p} & \text{if~$L$ is odd}\\ 
    \frac{(p+1)^{L/2+1}-2}{p} & \text{if~$L$ is even}
    \end{cases}.
\end{equation*}

\end{theorem}

\begin{remark}
According to Lemma~\ref{lemma:universalC1bound}, the prepare-and-shoot algorithm is strictly optimal in terms of~$C_1$.
In addition, since~$(p+1)^{L}<K$, by Lemma~\ref{lemma:universalC2bound} the algorithm is asymptotically optimal in terms of~$C_2$.
\end{remark}

\section{Draw and Loose: An Algorithm\\for Computing Vandermonde Matrices}\label{section:Vandermonde}

\ifFULL
In this section, we shift our attention to specific algorithms for computing Vandermonde matrices, for their prevalent use in Reed-Solomon codes.
That is, we tailor both the scheduling and the coding scheme for~$K\times K$ matrices~$A$ such that~$A_{i,j}=\alpha_{j}^i$ for~$i,j\in[0,K-1]$, where~$\alpha_0,\ldots,\alpha_{K-1}$ are distinct elements of~$\bbF_q$ (hence~$K\le q$). 
That is, every processor wishes to obtain~$\coded{x}_k=f(\alpha_k)$, an evaluation of the polynomial~$f(z)=\sum_{k\in[0,K-1]}x_kz^k$ at~$\alpha_k$.

Inspired by the Fast Fourier Transform algorithm, we show a method that computes the~\emph{Discrete Fourier Transform} (DFT) matrix (a special case of Vandermonde matrix) with the optimal~$C_1=C_2=\log_{p+1}K$ (see Remark~\ref{remark:bounds4SpecificMatrices}).
Later, this method serves as a primitive for computation of general Vandermonde matrices, and brings a significant gain in~$C_2$ compared with the universal algorithm described earlier.
\fi

\subsection{Computing a DFT Matrix}\label{section:DFT}
\begin{figure}
    \centering

\tikzset{every picture/.style={line width=0.75pt}} 

\begin{tikzpicture}[x=0.56pt,y=0.56pt,yscale=-1,xscale=1]

\draw  [fill={rgb, 255:red, 80; green, 227; blue, 194 }  ,fill opacity=1 ] (71,278) .. controls (71,273.58) and (74.58,270) .. (79,270) -- (103,270) .. controls (107.42,270) and (111,273.58) .. (111,278) -- (111,302) .. controls (111,306.42) and (107.42,310) .. (103,310) -- (79,310) .. controls (74.58,310) and (71,306.42) .. (71,302) -- cycle ;
\draw  [fill={rgb, 255:red, 80; green, 227; blue, 194 }  ,fill opacity=1 ] (111,278) .. controls (111,273.58) and (114.58,270) .. (119,270) -- (143,270) .. controls (147.42,270) and (151,273.58) .. (151,278) -- (151,302) .. controls (151,306.42) and (147.42,310) .. (143,310) -- (119,310) .. controls (114.58,310) and (111,306.42) .. (111,302) -- cycle ;
\draw  [fill={rgb, 255:red, 80; green, 227; blue, 194 }  ,fill opacity=1 ] (151,278) .. controls (151,273.58) and (154.58,270) .. (159,270) -- (183,270) .. controls (187.42,270) and (191,273.58) .. (191,278) -- (191,302) .. controls (191,306.42) and (187.42,310) .. (183,310) -- (159,310) .. controls (154.58,310) and (151,306.42) .. (151,302) -- cycle ;
\draw  [fill={rgb, 255:red, 80; green, 227; blue, 194 }  ,fill opacity=1 ] (191,278) .. controls (191,273.58) and (194.58,270) .. (199,270) -- (223,270) .. controls (227.42,270) and (231,273.58) .. (231,278) -- (231,302) .. controls (231,306.42) and (227.42,310) .. (223,310) -- (199,310) .. controls (194.58,310) and (191,306.42) .. (191,302) -- cycle ;
\draw  [fill={rgb, 255:red, 80; green, 227; blue, 194 }  ,fill opacity=1 ] (231,278) .. controls (231,273.58) and (234.58,270) .. (239,270) -- (263,270) .. controls (267.42,270) and (271,273.58) .. (271,278) -- (271,302) .. controls (271,306.42) and (267.42,310) .. (263,310) -- (239,310) .. controls (234.58,310) and (231,306.42) .. (231,302) -- cycle ;
\draw  [fill={rgb, 255:red, 80; green, 227; blue, 194 }  ,fill opacity=1 ] (271,278) .. controls (271,273.58) and (274.58,270) .. (279,270) -- (303,270) .. controls (307.42,270) and (311,273.58) .. (311,278) -- (311,302) .. controls (311,306.42) and (307.42,310) .. (303,310) -- (279,310) .. controls (274.58,310) and (271,306.42) .. (271,302) -- cycle ;
\draw  [fill={rgb, 255:red, 80; green, 227; blue, 194 }  ,fill opacity=1 ] (311,278) .. controls (311,273.58) and (314.58,270) .. (319,270) -- (343,270) .. controls (347.42,270) and (351,273.58) .. (351,278) -- (351,302) .. controls (351,306.42) and (347.42,310) .. (343,310) -- (319,310) .. controls (314.58,310) and (311,306.42) .. (311,302) -- cycle ;
\draw  [fill={rgb, 255:red, 80; green, 227; blue, 194 }  ,fill opacity=1 ] (351,278) .. controls (351,273.58) and (354.58,270) .. (359,270) -- (383,270) .. controls (387.42,270) and (391,273.58) .. (391,278) -- (391,302) .. controls (391,306.42) and (387.42,310) .. (383,310) -- (359,310) .. controls (354.58,310) and (351,306.42) .. (351,302) -- cycle ;
\draw  [fill={rgb, 255:red, 80; green, 227; blue, 194 }  ,fill opacity=1 ] (391,278) .. controls (391,273.58) and (394.58,270) .. (399,270) -- (423,270) .. controls (427.42,270) and (431,273.58) .. (431,278) -- (431,302) .. controls (431,306.42) and (427.42,310) .. (423,310) -- (399,310) .. controls (394.58,310) and (391,306.42) .. (391,302) -- cycle ;
\draw  [fill={rgb, 255:red, 80; green, 227; blue, 194 }  ,fill opacity=1 ] (70,188) .. controls (70,183.58) and (73.58,180) .. (78,180) -- (102,180) .. controls (106.42,180) and (110,183.58) .. (110,188) -- (110,212) .. controls (110,216.42) and (106.42,220) .. (102,220) -- (78,220) .. controls (73.58,220) and (70,216.42) .. (70,212) -- cycle ;
\draw  [fill={rgb, 255:red, 80; green, 227; blue, 194 }  ,fill opacity=1 ] (140,188) .. controls (140,183.58) and (143.58,180) .. (148,180) -- (172,180) .. controls (176.42,180) and (180,183.58) .. (180,188) -- (180,212) .. controls (180,216.42) and (176.42,220) .. (172,220) -- (148,220) .. controls (143.58,220) and (140,216.42) .. (140,212) -- cycle ;
\draw  [fill={rgb, 255:red, 80; green, 227; blue, 194 }  ,fill opacity=1 ] (210,188) .. controls (210,183.58) and (213.58,180) .. (218,180) -- (242,180) .. controls (246.42,180) and (250,183.58) .. (250,188) -- (250,212) .. controls (250,216.42) and (246.42,220) .. (242,220) -- (218,220) .. controls (213.58,220) and (210,216.42) .. (210,212) -- cycle ;
\draw  [fill={rgb, 255:red, 80; green, 227; blue, 194 }  ,fill opacity=1 ] (70,98) .. controls (70,93.58) and (73.58,90) .. (78,90) -- (102,90) .. controls (106.42,90) and (110,93.58) .. (110,98) -- (110,122) .. controls (110,126.42) and (106.42,130) .. (102,130) -- (78,130) .. controls (73.58,130) and (70,126.42) .. (70,122) -- cycle ;
\draw  [fill={rgb, 255:red, 74; green, 144; blue, 226 }  ,fill opacity=1 ] (440,98) .. controls (440,93.58) and (443.58,90) .. (448,90) -- (472,90) .. controls (476.42,90) and (480,93.58) .. (480,98) -- (480,122) .. controls (480,126.42) and (476.42,130) .. (472,130) -- (448,130) .. controls (443.58,130) and (440,126.42) .. (440,122) -- cycle ;
\draw  [fill={rgb, 255:red, 74; green, 144; blue, 226 }  ,fill opacity=1 ] (400,98) .. controls (400,93.58) and (403.58,90) .. (408,90) -- (432,90) .. controls (436.42,90) and (440,93.58) .. (440,98) -- (440,122) .. controls (440,126.42) and (436.42,130) .. (432,130) -- (408,130) .. controls (403.58,130) and (400,126.42) .. (400,122) -- cycle ;
\draw  [fill={rgb, 255:red, 74; green, 144; blue, 226 }  ,fill opacity=1 ] (360,98) .. controls (360,93.58) and (363.58,90) .. (368,90) -- (392,90) .. controls (396.42,90) and (400,93.58) .. (400,98) -- (400,122) .. controls (400,126.42) and (396.42,130) .. (392,130) -- (368,130) .. controls (363.58,130) and (360,126.42) .. (360,122) -- cycle ;
\draw  [fill={rgb, 255:red, 74; green, 144; blue, 226 }  ,fill opacity=1 ] (320,98) .. controls (320,93.58) and (323.58,90) .. (328,90) -- (352,90) .. controls (356.42,90) and (360,93.58) .. (360,98) -- (360,122) .. controls (360,126.42) and (356.42,130) .. (352,130) -- (328,130) .. controls (323.58,130) and (320,126.42) .. (320,122) -- cycle ;
\draw  [fill={rgb, 255:red, 74; green, 144; blue, 226 }  ,fill opacity=1 ] (280,98) .. controls (280,93.58) and (283.58,90) .. (288,90) -- (312,90) .. controls (316.42,90) and (320,93.58) .. (320,98) -- (320,122) .. controls (320,126.42) and (316.42,130) .. (312,130) -- (288,130) .. controls (283.58,130) and (280,126.42) .. (280,122) -- cycle ;
\draw  [fill={rgb, 255:red, 74; green, 144; blue, 226 }  ,fill opacity=1 ] (240,98) .. controls (240,93.58) and (243.58,90) .. (248,90) -- (272,90) .. controls (276.42,90) and (280,93.58) .. (280,98) -- (280,122) .. controls (280,126.42) and (276.42,130) .. (272,130) -- (248,130) .. controls (243.58,130) and (240,126.42) .. (240,122) -- cycle ;
\draw  [fill={rgb, 255:red, 74; green, 144; blue, 226 }  ,fill opacity=1 ] (200,98) .. controls (200,93.58) and (203.58,90) .. (208,90) -- (232,90) .. controls (236.42,90) and (240,93.58) .. (240,98) -- (240,122) .. controls (240,126.42) and (236.42,130) .. (232,130) -- (208,130) .. controls (203.58,130) and (200,126.42) .. (200,122) -- cycle ;
\draw  [fill={rgb, 255:red, 74; green, 144; blue, 226 }  ,fill opacity=1 ] (160,98) .. controls (160,93.58) and (163.58,90) .. (168,90) -- (192,90) .. controls (196.42,90) and (200,93.58) .. (200,98) -- (200,122) .. controls (200,126.42) and (196.42,130) .. (192,130) -- (168,130) .. controls (163.58,130) and (160,126.42) .. (160,122) -- cycle ;
\draw  [fill={rgb, 255:red, 74; green, 144; blue, 226 }  ,fill opacity=1 ] (120,98) .. controls (120,93.58) and (123.58,90) .. (128,90) -- (152,90) .. controls (156.42,90) and (160,93.58) .. (160,98) -- (160,122) .. controls (160,126.42) and (156.42,130) .. (152,130) -- (128,130) .. controls (123.58,130) and (120,126.42) .. (120,122) -- cycle ;
\draw    (90,130) .. controls (103.31,152.55) and (120.93,174.24) .. (137.64,186.35) ;
\draw [shift={(140,188)}, rotate = 213.96] [fill={rgb, 255:red, 0; green, 0; blue, 0 }  ][line width=0.08]  [draw opacity=0] (8.93,-4.29) -- (0,0) -- (8.93,4.29) -- cycle    ;
\draw  [fill={rgb, 255:red, 74; green, 144; blue, 226 }  ,fill opacity=1 ] (440,188) .. controls (440,183.58) and (443.58,180) .. (448,180) -- (472,180) .. controls (476.42,180) and (480,183.58) .. (480,188) -- (480,212) .. controls (480,216.42) and (476.42,220) .. (472,220) -- (448,220) .. controls (443.58,220) and (440,216.42) .. (440,212) -- cycle ;
\draw  [fill={rgb, 255:red, 74; green, 144; blue, 226 }  ,fill opacity=1 ] (370,188) .. controls (370,183.58) and (373.58,180) .. (378,180) -- (402,180) .. controls (406.42,180) and (410,183.58) .. (410,188) -- (410,212) .. controls (410,216.42) and (406.42,220) .. (402,220) -- (378,220) .. controls (373.58,220) and (370,216.42) .. (370,212) -- cycle ;
\draw  [fill={rgb, 255:red, 74; green, 144; blue, 226 }  ,fill opacity=1 ] (300.5,188) .. controls (300.5,183.58) and (304.08,180) .. (308.5,180) -- (332.5,180) .. controls (336.92,180) and (340.5,183.58) .. (340.5,188) -- (340.5,212) .. controls (340.5,216.42) and (336.92,220) .. (332.5,220) -- (308.5,220) .. controls (304.08,220) and (300.5,216.42) .. (300.5,212) -- cycle ;
\draw  [fill={rgb, 255:red, 74; green, 144; blue, 226 }  ,fill opacity=1 ] (440,278) .. controls (440,273.58) and (443.58,270) .. (448,270) -- (472,270) .. controls (476.42,270) and (480,273.58) .. (480,278) -- (480,302) .. controls (480,306.42) and (476.42,310) .. (472,310) -- (448,310) .. controls (443.58,310) and (440,306.42) .. (440,302) -- cycle ;
\draw    (78,130) -- (78,177) ;
\draw [shift={(78,180)}, rotate = 270] [fill={rgb, 255:red, 0; green, 0; blue, 0 }  ][line width=0.08]  [draw opacity=0] (8.93,-4.29) -- (0,0) -- (8.93,4.29) -- cycle    ;
\draw    (102,130) .. controls (128.72,145.96) and (165.11,168.69) .. (207.4,186.89) ;
\draw [shift={(210,188)}, rotate = 202.96] [fill={rgb, 255:red, 0; green, 0; blue, 0 }  ][line width=0.08]  [draw opacity=0] (8.93,-4.29) -- (0,0) -- (8.93,4.29) -- cycle    ;
\draw    (470,180) -- (469.03,133) ;
\draw [shift={(468.97,130)}, rotate = 88.81] [fill={rgb, 255:red, 0; green, 0; blue, 0 }  ][line width=0.08]  [draw opacity=0] (8.93,-4.29) -- (0,0) -- (8.93,4.29) -- cycle    ;
\draw    (458,180.01) .. controls (452.96,168.28) and (436.64,147.67) .. (422.05,132.16) ;
\draw [shift={(420,130)}, rotate = 46.17] [fill={rgb, 255:red, 0; green, 0; blue, 0 }  ][line width=0.08]  [draw opacity=0] (8.93,-4.29) -- (0,0) -- (8.93,4.29) -- cycle    ;
\draw    (440,188) .. controls (425.06,170.15) and (405.17,152.07) .. (382.15,131.88) ;
\draw [shift={(380,130)}, rotate = 41.2] [fill={rgb, 255:red, 0; green, 0; blue, 0 }  ][line width=0.08]  [draw opacity=0] (8.93,-4.29) -- (0,0) -- (8.93,4.29) -- cycle    ;
\draw    (390,180) .. controls (373.89,164.22) and (324.99,154.23) .. (302.05,132.09) ;
\draw [shift={(300,130)}, rotate = 47.12] [fill={rgb, 255:red, 0; green, 0; blue, 0 }  ][line width=0.08]  [draw opacity=0] (8.93,-4.29) -- (0,0) -- (8.93,4.29) -- cycle    ;
\draw    (378.5,180) .. controls (340.85,161.55) and (288.62,155.9) .. (262,131.88) ;
\draw [shift={(260,130)}, rotate = 44.65] [fill={rgb, 255:red, 0; green, 0; blue, 0 }  ][line width=0.08]  [draw opacity=0] (8.93,-4.29) -- (0,0) -- (8.93,4.29) -- cycle    ;
\draw    (402.5,180) .. controls (390.97,163.01) and (359.09,152.46) .. (341.82,132.24) ;
\draw [shift={(340,130)}, rotate = 52.34] [fill={rgb, 255:red, 0; green, 0; blue, 0 }  ][line width=0.08]  [draw opacity=0] (8.93,-4.29) -- (0,0) -- (8.93,4.29) -- cycle    ;
\draw    (308.5,180) .. controls (273.81,161.98) and (208.76,161.06) .. (181.61,131.83) ;
\draw [shift={(180,130)}, rotate = 49.95] [fill={rgb, 255:red, 0; green, 0; blue, 0 }  ][line width=0.08]  [draw opacity=0] (8.93,-4.29) -- (0,0) -- (8.93,4.29) -- cycle    ;
\draw    (300,188) .. controls (250.4,169.46) and (168.62,167.04) .. (143.46,132.05) ;
\draw [shift={(141.99,129.87)}, rotate = 57.89] [fill={rgb, 255:red, 0; green, 0; blue, 0 }  ][line width=0.08]  [draw opacity=0] (8.93,-4.29) -- (0,0) -- (8.93,4.29) -- cycle    ;
\draw    (332,180) .. controls (315.48,163.49) and (244.29,161.06) .. (221.64,132.27) ;
\draw [shift={(220,130)}, rotate = 56.35] [fill={rgb, 255:red, 0; green, 0; blue, 0 }  ][line width=0.08]  [draw opacity=0] (8.93,-4.29) -- (0,0) -- (8.93,4.29) -- cycle    ;
\draw    (460,270) .. controls (446.75,247.41) and (429.96,225.79) .. (413.35,213.65) ;
\draw [shift={(411,212)}, rotate = 34.1] [fill={rgb, 255:red, 0; green, 0; blue, 0 }  ][line width=0.08]  [draw opacity=0] (8.93,-4.29) -- (0,0) -- (8.93,4.29) -- cycle    ;
\draw    (472,270) -- (472,221.88) ;
\draw [shift={(472,218.88)}, rotate = 90] [fill={rgb, 255:red, 0; green, 0; blue, 0 }  ][line width=0.08]  [draw opacity=0] (8.93,-4.29) -- (0,0) -- (8.93,4.29) -- cycle    ;
\draw    (448,270) .. controls (421.32,253.98) and (384.87,230.03) .. (342.61,211.68) ;
\draw [shift={(340.02,210.56)}, rotate = 23.1] [fill={rgb, 255:red, 0; green, 0; blue, 0 }  ][line width=0.08]  [draw opacity=0] (8.93,-4.29) -- (0,0) -- (8.93,4.29) -- cycle    ;
\draw    (78,220) -- (78.94,267) ;
\draw [shift={(79,270)}, rotate = 268.85] [fill={rgb, 255:red, 0; green, 0; blue, 0 }  ][line width=0.08]  [draw opacity=0] (8.93,-4.29) -- (0,0) -- (8.93,4.29) -- cycle    ;
\draw    (90,220) .. controls (95.01,231.74) and (113.22,252.3) .. (127.94,267.84) ;
\draw [shift={(130,270)}, rotate = 226.31] [fill={rgb, 255:red, 0; green, 0; blue, 0 }  ][line width=0.08]  [draw opacity=0] (8.93,-4.29) -- (0,0) -- (8.93,4.29) -- cycle    ;
\draw    (110,212) .. controls (124.89,229.88) and (144.88,247.88) .. (167.86,268.12) ;
\draw [shift={(170,270)}, rotate = 221.34] [fill={rgb, 255:red, 0; green, 0; blue, 0 }  ][line width=0.08]  [draw opacity=0] (8.93,-4.29) -- (0,0) -- (8.93,4.29) -- cycle    ;
\draw    (160,220) .. controls (176.07,235.82) and (225.06,245.72) .. (247.95,267.91) ;
\draw [shift={(250,270)}, rotate = 227.26] [fill={rgb, 255:red, 0; green, 0; blue, 0 }  ][line width=0.08]  [draw opacity=0] (8.93,-4.29) -- (0,0) -- (8.93,4.29) -- cycle    ;
\draw    (172,220) .. controls (209.6,238.54) and (261.46,244.05) .. (288.01,268.11) ;
\draw [shift={(290,270)}, rotate = 224.79] [fill={rgb, 255:red, 0; green, 0; blue, 0 }  ][line width=0.08]  [draw opacity=0] (8.93,-4.29) -- (0,0) -- (8.93,4.29) -- cycle    ;
\draw    (148,220) .. controls (159.49,237.02) and (190.99,247.5) .. (208.19,267.76) ;
\draw [shift={(210,270)}, rotate = 232.48] [fill={rgb, 255:red, 0; green, 0; blue, 0 }  ][line width=0.08]  [draw opacity=0] (8.93,-4.29) -- (0,0) -- (8.93,4.29) -- cycle    ;
\draw    (242,220) .. controls (276.65,238.1) and (341.33,238.88) .. (368.39,268.17) ;
\draw [shift={(370,270)}, rotate = 230.09] [fill={rgb, 255:red, 0; green, 0; blue, 0 }  ][line width=0.08]  [draw opacity=0] (8.93,-4.29) -- (0,0) -- (8.93,4.29) -- cycle    ;
\draw    (250,212) .. controls (299.56,230.66) and (383.38,232.78) .. (408.53,267.81) ;
\draw [shift={(410,270)}, rotate = 238.03] [fill={rgb, 255:red, 0; green, 0; blue, 0 }  ][line width=0.08]  [draw opacity=0] (8.93,-4.29) -- (0,0) -- (8.93,4.29) -- cycle    ;
\draw    (218,220) .. controls (234.48,236.56) and (305.78,238.89) .. (328.36,267.73) ;
\draw [shift={(330,270)}, rotate = 236.49] [fill={rgb, 255:red, 0; green, 0; blue, 0 }  ][line width=0.08]  [draw opacity=0] (8.93,-4.29) -- (0,0) -- (8.93,4.29) -- cycle    ;

\draw (76.5,100) node [anchor=north west][inner sep=0.75pt]  [xscale=0.75,yscale=0.75] [align=left] {$\displaystyle f( z)$};
\draw (72,190) node [anchor=north west][inner sep=0.75pt]  [xscale=0.75,yscale=0.75] [align=left] {$\displaystyle f_{0}( z)$};
\draw (141.5,190) node [anchor=north west][inner sep=0.75pt]  [xscale=0.75,yscale=0.75] [align=left] {$\displaystyle f_{1}( z)$};
\draw (211,189) node [anchor=north west][inner sep=0.75pt]  [xscale=0.75,yscale=0.75] [align=left] {$\displaystyle f_{2}( z)$};
\draw (73,281) node [anchor=north west][inner sep=0.75pt]  [xscale=0.75,yscale=0.75] [align=left] {$\displaystyle f_{00}( z)$};
\draw (113,281) node [anchor=north west][inner sep=0.75pt]  [xscale=0.75,yscale=0.75] [align=left] {$\displaystyle f_{01}( z)$};
\draw (153,281) node [anchor=north west][inner sep=0.75pt]  [xscale=0.75,yscale=0.75] [align=left] {$\displaystyle f_{02}( z)$};
\draw (193,281) node [anchor=north west][inner sep=0.75pt]  [xscale=0.75,yscale=0.75] [align=left] {$\displaystyle f_{10}( z)$};
\draw (233,281) node [anchor=north west][inner sep=0.75pt]  [xscale=0.75,yscale=0.75] [align=left] {$\displaystyle f_{11}( z)$};
\draw (273,281) node [anchor=north west][inner sep=0.75pt]  [xscale=0.75,yscale=0.75] [align=left] {$\displaystyle f_{12}( z)$};
\draw (313,281) node [anchor=north west][inner sep=0.75pt]  [xscale=0.75,yscale=0.75] [align=left] {$\displaystyle f_{20}( z)$};
\draw (353,281) node [anchor=north west][inner sep=0.75pt]  [xscale=0.75,yscale=0.75] [align=left] {$\displaystyle f_{21}( z)$};
\draw (393,281) node [anchor=north west][inner sep=0.75pt]  [xscale=0.75,yscale=0.75] [align=left] {$\displaystyle f_{22}( z)$};
\draw (124,100) node [anchor=north west][inner sep=0.75pt]  [font=\large,xscale=0.75,yscale=0.75] [align=left] {$\displaystyle \gamma _{22}$};
\draw (310,190) node [anchor=north west][inner sep=0.75pt]  [font=\large,xscale=0.75,yscale=0.75] [align=left] {$\displaystyle \gamma _{2}$};
\draw (451,280) node [anchor=north west][inner sep=0.75pt]  [font=\large,xscale=0.75,yscale=0.75] [align=left] {$\displaystyle \gamma $};
\draw (380,190) node [anchor=north west][inner sep=0.75pt]  [font=\large,xscale=0.75,yscale=0.75] [align=left] {$\displaystyle \gamma _{1}$};
\draw (450,190) node [anchor=north west][inner sep=0.75pt]  [font=\large,xscale=0.75,yscale=0.75] [align=left] {$\displaystyle \gamma _{0}$};
\draw (164,100) node [anchor=north west][inner sep=0.75pt]  [font=\large,xscale=0.75,yscale=0.75] [align=left] {$\displaystyle \gamma _{12}$};
\draw (204,100) node [anchor=north west][inner sep=0.75pt]  [font=\large,xscale=0.75,yscale=0.75] [align=left] {$\displaystyle \gamma _{02}$};
\draw (244,100) node [anchor=north west][inner sep=0.75pt]  [font=\large,xscale=0.75,yscale=0.75] [align=left] {$\displaystyle \gamma _{21}$};
\draw (284,100) node [anchor=north west][inner sep=0.75pt]  [font=\large,xscale=0.75,yscale=0.75] [align=left] {$\displaystyle \gamma _{11}$};
\draw (324,100) node [anchor=north west][inner sep=0.75pt]  [font=\large,xscale=0.75,yscale=0.75] [align=left] {$\displaystyle \gamma _{01}$};
\draw (364,100) node [anchor=north west][inner sep=0.75pt]  [font=\large,xscale=0.75,yscale=0.75] [align=left] {$\displaystyle \gamma _{20}$};
\draw (404,100) node [anchor=north west][inner sep=0.75pt]  [font=\large,xscale=0.75,yscale=0.75] [align=left] {$\displaystyle \gamma _{10}$};
\draw (444,100) node [anchor=north west][inner sep=0.75pt]  [font=\large,xscale=0.75,yscale=0.75] [align=left] {$\displaystyle \gamma _{00}$};

\end{tikzpicture}

    \caption{Illustration of the trees with~$K=9$ and~$p=2$. (left) The polynomial tree rooted at $f(z)=x_0+x_1z+\cdots+x_{8}z^{8}$. 
    The polynomials in the first level are 
    $f_0(z)=x_0+x_3z+x_{6}z^{3}$,
    $f_1(z)=x_1+x_4z+x_{7}z^{3}$, etc.  
    see~\eqref{eq:treePolynomialRecursion}.
    (right) The tree of field elements rooted at~$\gamma=1$, with~$\gamma_{1}=g^3,\gamma_{10}=g$, etc, where~$g$ is a generator of~$\bbF_q$, see~\eqref{eq:treeNodeGamma}. }
    \label{fig:tree}
\end{figure}
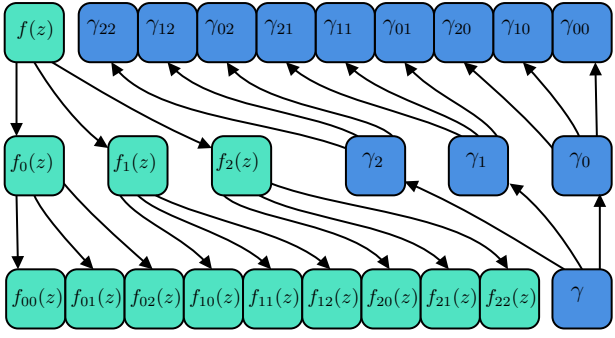

\ifFULL
Assume that~$K=(p+1)^H$ for some positive integer~$H$, that~$K\vert q-1$, and let~$\beta=g^\frac{q-1}{K}$ be a primitive~$K$-th root of unity, where~$g$ is a generator of~$\bbF_q$.
A Discrete Fourier Transform (DFT) matrix~$D_K$ is a Vandermonde one with~$\alpha_k=\beta^k$, i.e.,
\begin{equation}\label{eq:DFTmatrix}
D_K=\begin{bmatrix}
    1      &1                &1              &\cdots & 1                 \\
    1      &\beta            &\beta^2        &\cdots & \beta^{K-1}       \\
    \vdots &\vdots           &\vdots         &\ddots & \vdots            \\
    1      &\beta^{(K-1)}   &\beta^{2(K-1)} &\cdots & \beta^{(k-1)(K-1)}    \\
    \end{bmatrix}.
\end{equation}
\fi 
The proposed method relies on two complete~$(p+1)$-ary trees, a tree of field elements and a tree of polynomials. 
Each tree is of height~$H$, and there are~$(p+1)^h$ nodes at level~$h\in[0,H]$.
Therefore, a node at level~$h$ can be represented by~$h$ digits using $(p+1)$-radix.

The tree of field elements is defined as follows. A node at level~$h\in[0,H]$ is identified by~$h$ digits $k_{h-1}\ldots k_0$ in~$[0,p]$ ($k_{h-1}$ being the most significant), and contains the element \begin{equation}\label{eq:treeNodeGamma}
\gamma_{k_{h-1}\cdots k_{0} } \triangleq(\beta^{k_{h-1}(p+1)^{h-1}+\cdots+k_{0}})^{(p+1)^{H-h}}.    
\end{equation}
It is readily verified that for each of the leaves (at level~$H$), we have that~$
\gamma_{k_{H-1}\cdots k_{0}}=\beta^{k}$ is the evaluation point of the processor indexed by~$k={k_{H-1}(p+1)^{H-1}+\cdots+k_{0}}$. For the root (at level~$0$), which is represented by~$0$ digits, we have~$\gamma=1$. 
Indices of sibling nodes differ only in the most significant (leftmost) digit, and index of a parent node is given by omitting the most significant digit of its child.
That is, nodes~$\gamma_{0k_{h-1}\cdots k_{0}},\ldots,\gamma_{pk_{h-1}\cdots k_{0}}$ are the children of the same parent node~$\gamma_{k_{h-1}\cdots k_{0}}$. By Equation~\eqref{eq:treeNodeGamma}, every child is a distinct~$(p+1)$-th root of its parent, i.e., for every~$\rho\in[0,p]$,
\ifFULL\begin{align}
    (\gamma&_{\rho k_{h-1}\cdots k_{0}})^{p+1}\nonumber\\
    =\;& ((\beta^{\rho(p+1)^{h}+k_{h-1}(p+1)^{h-1}+\cdots+k_{0}})^{(p+1)^{H-h-1}})^{p+1}\nonumber\\
    =\;& (\beta^{\rho(p+1)^{h}+k_{h-1}(p+1)^{h-1}+\cdots+k_{0}})^{(p+1)^{H-h}}\nonumber\\
    =\;& (\beta^{k_{h-1}(p+1)^{h-1}+\cdots+k_{0}})^{(p+1)^{H-h}}\cdot \beta^{\rho(p+1)^H}\nonumber\\
    =\;& \gamma_{k_{h-1}\cdots k_{0}} \cdot (g^\frac{q-1}{K})^{\rho K}=\gamma_{k_{h-1}\cdots k_{0}}.
\end{align}\fi

The tree of polynomials is defined recursively from the root labelled by~$f(z)$. 
For a non-leaf node labelled by~$
f_{k_{0}\cdots k_{h-1}}(z)=b_0z^0+\cdots+b_{D-1}z^{D-1}
$ at level~$h$, its~$\rho$-th child node is defined as
\begin{align}\label{eq:treePolynomialRecursion}
f_{ k_{0}\cdots k_{h-1}\rho}(z)= \sum_{d=\rho\bmod(p+1)} b_dz^\frac{d-\rho}{p+1}.
\end{align}
Therefore, every non-leaf node is a combination of its children evaluated at~$z^{p+1}$, i.e.,
\begin{equation}\label{eq:combinationOfChildren}
    f_{k_{0}\cdots k_{h-1}}(z)=\sum_{\rho\in[0,p]}z^\rho f_{k_{0}\cdots k_{h-1}\rho}(z^{p+1}).
\end{equation}
Note that the leaf node~$f_{k_{0},\ldots,k_{H-1}}=x_{k'}$ is the initial packet of the processor indexed by~$k'=k_{H-1}+\cdots+k_0(p+1)^{H-1}$.
That is, the~$(p+1)$-radix representations of~$k={k_{H-1}(p+1)^{H-1}+\cdots+k_{0}}$ (defined above) and~${k'}$ have reversed order.
See Figure~\ref{fig:tree} for illustrations of both trees.

The proposed algorithm is defined recursively using the above trees. 
Define~$Q(k,t)=f_{k_{H-1}\cdots k_{t}}(\gamma_{k_{t-1}\cdots k_{0}})$,
and hence every processor~$k$ initially has~$Q(k,0)=f_{k_{H-1}\cdots k_{0}}(\gamma)= f_{k'_{0}\cdots k'_{H-1}}(\gamma)=x_k$
at the beginning of the algorithm, and wishes to obtain~$Q(k,H)=f(\gamma_{k_{H-1}\cdots k_{0}})=f(\beta^k)=\coded{x}_k$.

Assume that processor~$k$ has~$Q(k,t)$ after round~$t$,
and we show how it can obtain~$Q(k,t+1)$ in one round.
By Equation~\eqref{eq:combinationOfChildren} and Equation~\eqref{eq:treeNodeGamma} we have
\ifFULL\begin{align}\label{eq:linearCombination}
   Q(k,t+1)&=f_{k_{H-1}\cdots k_{t+1}}(\gamma_{k_{t}\cdots k_{0}}) \nonumber\\
    &=\sum_{\rho\in[0,p]}(\gamma_{k_{t}\cdots k_{0}})^\rho f_{k_{H-1}\cdots k_{t+1}\rho}((\gamma_{k_{t}\cdots k_{0}})^{p+1})\nonumber\\
    &=\sum_{\rho\in[0,p]}(\gamma_{k_{t}\cdots k_{0}})^\rho f_{k_{H-1}\cdots k_{t+1}\rho}(\gamma_{k_{t-1}\cdots k_{0}})\nonumber\\
   &=\sum_{\rho\in[0,p]}(\gamma_{k_{t}\cdots k_{0}})^\rho Q(k^{(t)}_\rho,t),
\end{align}\fi
where~$k^{(t)}_\rho$ is 
represented by~$k_{H-1}\cdots k_{t+1}\rho k_{t-1}\cdots k_0$ in~$(p+1)$-radix, i.e., it differs from~$k$ only in the~$t$'th digit.
Therefore, the desired~$Q(k,t+1)$ is a linear combination of~$Q(k^{(t)}_0,t),\ldots,Q(k^{(t)}_p,t)$.
Written in matrix form 
we have
\begin{equation}\label{eq:step}
    \begin{bmatrix}
    Q(k_0^{(t)},t+1)\\
    \vdots\\
    Q(k^{(t)}_p,t+1)
    \end{bmatrix}=
    A_k^{(t)}\cdot
    \begin{bmatrix}
    Q(k_0^{(t)},t)\\
    \vdots\\
    Q(k^{(t)}_p,t)
    \end{bmatrix},
\end{equation}
\begin{equation}\label{eq:linearCombinationMatrix}
\text{where}~
    A_k^{(t)}=\begin{bmatrix}
    \gamma_{(0k_{t-1}\cdots k_{0}})^0 & \cdots &  (\gamma_{0k_{t-1}\cdots k_{0}})^P \\
    \vdots                        & \ddots & \vdots\\
    \gamma_{(pk_{t-1}\cdots k_{0}})^0 & \cdots &  (\gamma_{pk_{t-1}\cdots k_{0}})^P \\
    \end{bmatrix}.
\end{equation}

At round~$t+1$ of the proposed algorithm, every processor~$k$ broadcasts~$Q(k,t)$ to the~$p$ processors having the same index except for the~$t$-th digit. 
In the same round, processor~$k$ receives $Q(k_\rho^{(t)},t)$ for every~$\rho\in[0,p]$ from these processors (including itself). 
Then, processor~$k$ obtains~$Q(k,t+1)$ by linearly combining the received packets based on Equation~\eqref{eq:linearCombination}.

Recall that every processor~$k$ has~$Q(k,0)$ at beginning, it obtains the coded packet~$\coded{x}_k=Q(k,H)$ after~$H$ rounds by repeating the above operation~$H$ times. 
Since exactly one packet is transmitted through each of the~$p$ ports during each operation, we have the following theorem.
\begin{theorem}\label{theorem:optimalityOfDrawAndLoose}
    The above algorithm for computing a DFT matrix has~$C_1=C_2=H=\log_{p+1}K$, which is strictly optimal.
\end{theorem}

\begin{remark}
As shown in Remark~\ref{remark:bounds4SpecificMatrices}, the proposed algorithm has the strictly optimal~$C_1$ value. 
Further, the~$C_2$ value is also optimal, since during each round only 1 packet is communicated through each port.
\ifFULL This is an exponential improvement over the universal algorithm. \fi
\end{remark}

Next, we emphasize the~\emph{invertibility} of the presented algorithm in the following lemma; this will be useful in the sequel.
\begin{lemma}\label{lemma:invertibilityOfLoose}
    The above algorithm can be used to compute the inverse of a DFT matrix, with the same~$C_1$ and~$C_2$.
\end{lemma}
\ifFULL \begin{proof}
Since the matrix~$A_k^{(t)}$ defined in Equation~\eqref{eq:linearCombinationMatrix} is an invertible Vandermonde matrix, it follows that each step of the induction is invertible.
That is, the processor~$k$ in possession of~$Q(k,t+1)$ can obtain~$Q(k,t)$ in one round of communication, sending and receiving~$1$ packet through each of its ports, for every~$t\in[0,H]$.
Therefore, a processor~$k$ in possession of~$Q(k,H)$ can obtain~$Q(k,0)$ in~$H$ rounds, with the optimal~$C_1$ and~$C_2$ as shown in Theorem~\ref{theorem:optimalityOfDrawAndLoose}.
\end{proof} \fi

\subsection{Generalization}

The aforementioned algorithm computes a unique Vandermonde matrix with strictly optimal~$C_1$ and~$C_2$, but requires that~$K=(p+1)^H$ for some~$H$ and that~$K\mid q-1$. 
In cases where~$K\nmid q-1$ and~$K\le q-1$, let~$H$ be the maximum integer such that~$(p+1)^H$ divides~$\gcd(K,q-1)$, and denote~$K=M\cdot(p+1)^H$. We use the above DFT algorithm as a primitive for improved~$C_2$ with respect to prepare-and-shoot in the computation of multiple other Vandermonde matrices.

Let~$Z=(p+1)^H$, and denote processor~$P_{i,j}=j+Z\cdot i$ by two indices~$j\in[0,Z-1]$ and~$i\in[0,M-1]$.
We define the evaluation point of processor~$P_{i,j}$ as~$\alpha_{i,j}= \alpha_i\cdot\beta_j$, where~$\alpha_i=g^{\varphi(i)}$ and~$\beta_j=  g^{j\cdot\frac{q-1}{Z}}$, with~$g$ being a generator of~$\bbF_q$ and~$\varphi$ being any injective map from~$[0,M-1]$ to~$[0,(q-1)/Z-1]$, which exists since~$q-1\ge ZM=K$. 
\ifFULL
These definitions guarantee that no two evaluation points are identical.
\else
\fi
In addition, since there exists~$\binom{(q-1)/Z}{M}$ possible choices for~$\varphi$, it follows that the proposed algorithm computes this many different Vandermonde matrices up to permutation of columns.

Recall that the coded packet~$\coded{x}_{i,j}$ desired by processor~$P_{i,j}$ is an evaluation of~$f(z)$ at~$\alpha_{i,j}$.
Moreover, we have
\ifFULL\begin{equation}\label{eq:extractEll}
    \begin{split}
f(\alpha_{i,j})&=\sum_{k=0}^{K-1}x_k(\alpha_{i,j})^k=\sum_{k=0}^{K-1}x_k\alpha_i^k\beta_j^k\\
&=\sum_{\ell=0}^{Z-1}\beta_j^\ell\sum_{k=\ell\bmod Z}x_k\alpha_{i}^k\beta_{j}^{k-\ell}. 
    \end{split}
\end{equation}
\fi
Since~$g^{q-1}=1$, it follows that~$\beta_j^{k-\ell}=g^{j\cdot\frac{q-1}{Z}(k-\ell)}=1$ whenever~$k=\ell\bmod Z$, and hence
\begin{equation}
        \coded{x}_{i,j}\overset{\eqref{eq:extractEll}}{=}\sum_{\ell=0}^{Z-1}\beta_j^\ell\sum_{k=\ell\bmod Z}x_k\alpha_{i}^k=\sum_{\ell=0}^{Z-1}\beta_j^\ell f_\ell(\alpha_i),
\end{equation}
where~$f_\ell(z)=\sum_{w=0}^{M-1}x_{w,\ell}z^{\ell+Z\cdot w}$. 
In matrix form,
\begin{equation}\label{eq:loosePhase}
\coded{X}\triangleq
\begin{bmatrix}
\coded{x}_{0,0} & \cdots &\coded{x}_{0,Z-1}\\
\vdots & \ddots &\vdots\\
\coded{x}_{M-1,0} & \cdots &\coded{x}_{M-1,Z-1}\\
\end{bmatrix}=F \cdot D_L,
\end{equation}
where~$D_L$ is an~$Z\times Z$ DFT matrix, 
and
\ifFULL
\begin{align}\label{eq:drawPhase}
F =\begin{bmatrix}
    f_0(\alpha_0) &\cdots & f_{Z-1}(\alpha_0) \\
    \vdots        &\ddots & \vdots \\
    f_0(\alpha_{M-1}) &\cdots &f_{Z-1}(\alpha_{M-1})\\
    \end{bmatrix}.
\end{align}
\fi
Notice that the~$j$-th column of~$F$ satisfies
\begin{align}\label{equation:Fjthcolumn}
&[f_j(\alpha_0),\cdots,f_j(\alpha_{M-1})]^\intercal=\\\nonumber
    &\begin{bmatrix}
    \alpha_0^j & & \\
    & \ddots & \\
    & & \alpha_{M-1}^j
  \end{bmatrix}\cdot
    \begin{bmatrix}
    \alpha_0^{Z\cdot0}      & \cdots &  \alpha_0^{Z\cdot(M-1)}\\
    \vdots          & \ddots & \vdots \\
    \alpha_{M-1}^{Z\cdot0}  & \cdots & \alpha_{M-1}^{Z\cdot(M-1)} 
    \end{bmatrix}\cdot
    \begin{bmatrix}
    {x}_{0,j} \\
    \vdots  \\
    {x}_{M-1,j}
    \end{bmatrix}
\end{align}
The protocol proceeds in two phases. 

\textbf{Draw Phase:}
The objective of this phase is for every processor~$P_{i,j}$ to obtain~$f_j(\alpha_i)$.
As shown above~\eqref{equation:Fjthcolumn},~$f_j(\alpha_0),\ldots,f_j(\alpha_{M-1})$ are given by multiplying $[x_{0,j},\ldots,x_{M-1,j}]^\intercal$ by a Vandermonde matrix, denoted by~$V$, and a diagonal matrix $\diag(\alpha_0^j,\cdots,\alpha_{M-1}^j)$.
Therefore, this problem can be solved in parallel by~$Z$ all-to-all encode operations. For every~$j\in[0,Z-1]$, processors~$P_{0,j},\ldots,P_{M-1,j}$ collectively compute the matrix~$V$ using prepare-and-shoot (Section~\ref{section:universalAlgorithm}).
Once completed, every processor~$P_{i,j}$ locally multiplies the resulting packet with~$\alpha_i^j$ and obtains~$f_j(\alpha_i)$\ifFULL
, which is the element in the~$i$-th row and~$j$-th column of the matrix~$F$.
\fi

\textbf{Loose Phase:}
As shown in Equation~\eqref{eq:loosePhase}, the coded packets~$\coded{x}_{i,0},\ldots,\coded{x}_{i,Z-1}$ are linear combinations, defined by the DFT matrix~$D_L$, of the elements~$f_0(a_i),\ldots,f_{Z-1}(a_i)$ in the~$i$-th row of the matrix~$F$. For every~$j\in[0,Z-1]$, processors~$P_{i,0},\ldots,P_{i,Z-1}$ collectively compute~$D_L$ using the specialized algorithm a for DFT matrix (Section~\ref{section:DFT}). After~$H$ rounds, every processor~$P_{i,j}$ obtains the coded packet~$\coded{x}_{i,j}$.


Let~$\Psi(M)$ be the~$C_2$ in the prepare-and-shoot algorithm to compute any~$M\times M$ matrix.
Observe that the draw phase takes $C_1=\ceil{\log_{p+1}M}$ rounds and~$C_2=\Psi(M)$ communication, and the loose phase takes~$C_1=H$ rounds and~$C_2=H$ communication. 
Therefore, we have the following.

\begin{theorem}\label{theorem:Vandermonde}
The draw-and-loose algorithm can compute~$\binom{(q-1)/Z}{M}$ different Vandermonde matrices (up to permutation of columns) with $C_1=\ceil{\log_{p+1}K}$ and~$C_2=H+\Psi(M)$. 
In particular, if~$M\le p+1$ then~$\Psi(M)=1$, i.e.,~$C_1=C_2=\ceil{\log_{p+1}K}$.
\end{theorem}
\begin{remark}
    Note that draw-and-loose can compute any Vandermonde matrix. 
    Yet, significant gains w.r.t prepare-and-shoot will be possible in cases where~$H$ is large.
    \ifFULL In particular, in cases where~$H=0$ the draw-and-loose algorithm does not provide gains over the universal prepare-and-shoot.\fi
\end{remark}


\begin{lemma}\label{lemma:invertibilityOfDrawLoose}
    Similar to Lemma~\ref{lemma:invertibilityOfLoose}, the above algorithm can be used to compute the inverse of a Vandermonde matrix, with the same~$C_1$ and~$C_2$.
\end{lemma}
\ifFULL \begin{proof}
The invertibility of the loose phase is given in Lemma~\ref{lemma:invertibilityOfLoose}.
In the draw phase, since the Vandermonde matrix~$V$ defined in Equation~\eqref{eq:drawPhase} is invertible, this step can be inverted by computing the inverse of~$V$ using prepare-and-shoot. 

Together, the inverse of a Vandermonde matrix can be computed by first inversing the loose phase, and computing the inverse of~$V$ with prepare-and-shoot.
\end{proof} \fi

\section{Computing Lagrange Matrices\\with Invertible Draw-and-Loose}
Lagrange matrices were recently popularized for their use in coded computing~\cite{LCC}. For sets~$\{\alpha_i\}_{i=1}^K$and~$\{\omega_i\}_{i=1}^K$, each with~$K$ distinct elements in~$\bbF_q$, let 
\ifFULL
\begin{align*}
A=
\begin{bmatrix}
\Phi_1(\alpha_1)  & \ldots &\Phi_1(\alpha_K) \\
\vdots &\ddots&\vdots\\
\Phi_{K}(\alpha_1) &  \ldots & \Phi_{K}(\alpha_K) \\
\end{bmatrix},     \Phi_k(z)= \prod_{j\neq k} \frac{z-\omega_j}{\omega_k - \omega_j}.
\end{align*}
\fi
In this section we sketch an extension of draw-and-loose which computes Lagrange matrices.
Evidently, computing a Lagrange matrix can be described as follows.
Every processor~$k$ has~$x_k=f(\omega_k)$; they together form a point-value representation of a polynomial~$f(z)=\sum_{k\in[0,K-1]}f_kz^k$ of degree~$K-1$ at~$\omega_0,\ldots,\omega_{K-1}$.
Every processor~$k$ wants~$\coded{x}_k=f(\alpha_k)$, i.e., another point-value representation of~$f(z)$ on 
~$\alpha_0,\ldots,\alpha_{K-1}$.

Therefore, computing a Lagrange matrix is possible by two consecutive computations. First, compute the \textit{inverse} of a Vandermonde matrix~$V(\omega_1,\ldots,\omega_K)$ (Lemma~\ref{lemma:invertibilityOfDrawLoose}) in order to obtain the coefficients of the polynomial~$f$. 
Second, compute the Vandermonde matrix~$V(\alpha_1,\ldots,\alpha_K)$ in order to evaluate~$f$ at~$\alpha_k$ for every~$k\in[K]$. 
This yields the following, in which~$C_i(x)$ is the~$C_i$ measure of draw-and-loose over~$V(x_1,\ldots,x_k)$, for~$i\in\{1,2\}$ and~$x\in\{\alpha,\omega\}$.
\begin{theorem}
    The above algorithm computes a Lagrange matrix with~$C_1=C_1(\omega)+C_1(\alpha)$ and~$C_2=C_2(\omega)+C_2(\alpha)$.
\end{theorem}
\ifFULL \begin{proof}

Written in matrix form, the initial packets
\begin{equation}
(x_0,\ldots,x_{K-1})=(f_0,\ldots,f_{K-1})\begin{bmatrix}
    \omega_0^0      & \cdots & \omega_{n-1}^0 \\
    \vdots          & \ddots & \vdots \\
    \omega_0^{n-1}  & \cdots & \omega_{n-1}^{n-1} 
    \end{bmatrix}
\end{equation}
are linear combination of~$f_0,\ldots,f_{K-1}$ defined by the Vandermonde matrix~$V(\omega_0,\ldots,\omega_{K-1})$. 
By the invertibility of draw-and-loose (lemma~\ref{lemma:invertibilityOfDrawLoose}), this step can be completed with the same~$C_1$ and~$C_2$ introduced in Theorem~\ref{theorem:Vandermonde}, and every processor~$k$ obtains~$f_k$.
Next, the coded packets
\begin{equation}
(\coded{x}_0,\ldots,\coded{x}_{K-1})=(f_0,\ldots,f_{K-1})\begin{bmatrix}
    \alpha_0^0      & \cdots & \alpha_{n-1}^0 \\
    \vdots          & \ddots & \vdots \\
    \alpha_0^{n-1}  & \cdots & \alpha_{n-1}^{n-1} 
    \end{bmatrix},
\end{equation}
are linear combination of~$f_0,\ldots,f_{K-1}$ defined by the Vandermonde matrix~$V(\alpha_0,\ldots,\alpha_{K-1})$. 
Since every processor~$k$ has obtained~$f_k$, this phase can be completed using Vandermonde algorithm.\qedhere
\end{proof} \fi

\section{Discussion and future work}
Directions for future work include extending the results to other specific matrices, e.g., Cauchy matrices and Moore matrices, and to further study algorithms and lower bounds for Vandermonde and Lagrange matrices.
Also, incorporating computation and storage constraints of processors is an interesting direction for future research.

\end{document}